\documentclass[aps,prd,twocolumn,reprint,showpacs,groupedaddress]{revtex4-1}

\usepackage{multirow}
\usepackage{graphicx}

\begin{document}

\title{Influence of hadronic interaction models and the cosmic-ray spectrum on the high-energy atmospheric muon and neutrino flux}

\author{Anatoli Fedynitch}
\author{Julia Becker Tjus}
\affiliation{Ruhr-Universit\"at Bochum, Fakult\"at f\"ur Physik \& Astronomie, Theoretische Physik IV, D-44780 Bochum, Germany}

\author{Paolo Desiati}
\affiliation{Wisconsin IceCube Particle Astrophysics Center (WIPAC) and Department of Astronomy, University of Wisconsin, Madison, WI 53706, U.S.A.}

\date{\today}

\begin{abstract}
The recent observations of muon charge ratio up to about 10 TeV and of
atmospheric neutrinos up to energies of about 400 TeV has triggered a
renewed interest into the high-energy interaction models and cosmic
ray primary composition. A reviewed calculation of lepton spectra
produced in cosmic-ray induced extensive air showers is carried out
with a primary cosmic-ray spectrum that fits the latest direct
measurements below the knee. In order to achieve this, we used a full
Monte Carlo method to derive the inclusive differential spectra
(yields) of muons, muon neutrinos and electron neutrinos at the
surface for energies between 80 GeV and hundreds of PeV. Using these
results the differential flux and the flavor ratios of leptons were
calculated. The air shower simulator {\sc corsika} 6.990 was used for
showering and propagation of the secondary particles through the
atmosphere, employing the established high-energy hadronic interaction
models {\sc sibyll} 2.1, {\sc qgsjet-01} and {\sc qgsjet-ii-03}. We show that the
performance of the interaction models allows makes it possible to
predict the spectra within experimental uncertainties, while {\sc sibyll}
generally yields a higher flux at the surface than the qgsjet
models. The calculation of the flavor and charge ratios has lead to
inconsistent results, mainly influenced by the different
representations of the K/$\pi$ ratio within the models. The influence
of the knee of cosmic-rays is reflected in the secondary spectra at
energies between 100 and 200 TeV. Furthermore, we could quantify
systematic uncertainties of atmospheric muon- and neutrino fluxes,
associated to the models of the primary cosmic-ray spectrum and the
interaction models. For most recent parametrizations of the cosmic-ray 
primary spectrum, atmospheric muons can be determined with an
uncertainty smaller than $^{+15}_{-13}$\% of the average
flux. Uncertainties of the muon- and electron neutrino fluxes can be calculated within an average error of
$^{+32}_{-22}$\% and $^{+25}_{-19}$\%, respectively. 
\end{abstract}

\pacs{}
\keywords{atmospheric neutrinos, atmospheric muons, prompt, muon charge ratio, cosmic-rays, particle physics}

\maketitle

\section{Introduction}
\label{sec:intro}
After a century from the discovery of cosmic-rays, their origin is still the major quest in astrophysics. If ions are efficiently accelerated in diffusive shocks, supernova remnants in our Galaxy could be the major source of cosmic-rays up to about $10^{17}$ eV. In general, if hadronic particles are accelerated, a fraction of them must interact within their sources or in nearby molecular clouds to produce mesons, which eventually decay into high-energy gamma rays and neutrinos with the energy spectrum $\sim E^{-2}$ of the accelerated cosmic-rays, while the rest propagates across the Galaxy until their detection on Earth. The observed all-particle cosmic-ray spectrum can be generally described as a power law $E^{-\gamma}$, with $\gamma \sim 2.7$ up to $\sim 3 \times 10^{15} ~ \rm{eV}$ (the so-called knee) and with $\gamma \sim 3.0$ up to about $10^{17}$ eV, above which energy, cosmic-rays are believed to be of extragalactic origin. The interaction of these cosmic-rays in the dense Earth's atmosphere produce mesons and therefore, muons and neutrinos with a steep spectrum of $\sim E^{-3.7}$ \cite{gaisser_book}. The search for fluxes of extra-terrestrial neutrinos rely on the fact that they should dominate at energy in excess of a few hundreds TeV over the atmospheric neutrino background. On the other hand, at this energy range the atmospheric neutrinos are mostly susceptible to large uncertainties due to cosmic-ray spectral shape and composition above the knee, and to details of hadronic interaction models for the production of mesons in the atmosphere. A good understanding of the atmospheric neutrino flux is important for the identification of high-energy cosmic neutrinos and therefore of the origin of the cosmic-rays.

The lepton fluxes in the atmosphere are produced in the hadronic interactions of the cosmic-ray nucleons with air nuclei. The most abundant species of the short lived intermediate particles produced is the pion as the lightest known meson, followed by heavier particles with shorter life times such as kaons, $D$-mesons etc. The production ratio of particles to anti-particles incorporates the isospin symmetry, therefore the inclusive $\pi^+$-spectrum from protons equals to the $\pi^-$-spectrum of neutrons. The regeneration and flavor changing processes, in the language of z-factors these are e.g. $Z_{pp}$ and $Z_{pn}$, influence the type of nucleon interactions as a function of the atmospheric depth $X$. The charge ratio of kaons is more complex, since charged and neutral kaons are involved in the production of leptons. The isospin symmetries are different because the associated production of $K+$ via $p + A \to \Lambda + K^+ + anything$ is not compensated by an appropriate channel for $K^-$, but instead $n + A \to K^0 + anything$. Therefore, the charge ratio of the mesons participating in the lepton flux at the surface also depends on the species of mother meson. During the development of the cascade, the mesons travel through a medium with an increasing desity, thus the chance for an decay prior having an interaction with air is suppressed as the meson energy increases. The particle energy at which the probabilities for interaction and decay are balanced is called critical energy. The approxmate values for a vertical transversal of the atmosphere are $\epsilon_\pi \approx 115$ GeV for pions, $\epsilon_{K^\pm} \approx 850$ GeV for charged kaons, $\epsilon_{K^0_L} \approx 210$ GeV for neutral (long) kaons and $\epsilon_{ch} > 10^7$ GeV for particles containing charmed quarks \cite{thunman_1996}. For inclined cascades these critical energies have to be multiplied by $1/\cos(\theta)$, where $\theta$ is the zentih angle, to represent the elongated air density profile the meson sees on it's path towards the surface. 

The combination of the above mentioned effects of the meson species, the life time and the belonging to a certain isospin multiplet determines the observed increase of $\mu^+/\mu^-$ ratio in the TeV range \cite{minos_mucharge,opera_muon_charge,interpretation_muon_charge}. The role of charged kaons is more important for neutrinos, since their energy is almost equally split between the $\mu$ and the $\nu_\mu$ in the leptonic decay, while the pions' energy is mostly carried by the muons. Due to the kaons' shorter life time, they consitute the dominant  source of neutrinos above a few hundreds GeV. The K/$\pi$ ratio also determines the response of muons and neutrinos to the seasonal variations in the stratospheric temperatures, which can be used to probe the contributions of heavy mesons in the extensive air showers \cite{Tilav:2010vy,desiati_2010,minos_atmospheric, minos_2011}. Above a few 100 TeV the decay of heavy charmed mesons is expected to contribute to the lepton spectrum. Due to large uncertainties in charm production at large Feynman $x$, it is not known where such transition actually occurs, and it is model dependent.

The atmospheric flux of $\nu_\mu + \overline{\nu}_\mu$ has been measured in a wide energy range from 1 GeV to a few hundreds TeV. The IceCube Observatory reported the first determination of the high-energy neutrino flux in the hundreds TeV region \cite{icecube_ic40_2010,amanda_2010}. In order to calculate the atmospheric neutrino intensities precisely, we need detailed information about (i) the primary cosmic-ray spectra at the top of the atmosphere, (ii) the hadronic interactions between cosmic-rays and atmospheric nuclei, (iii) the propagation of cosmic-ray particles inside the atmosphere, and (iv) the decay of the secondary particles. The comparisons between various calculations and with direct measurements makes it possible to assess how the contribution of experimental uncertainties in the primary spectrum and in the different hadronic models affects the atmospheric neutrino spectrum at the energy range that is relevant to the current neutrino telescopes. In this paper, the Monte Carlo calculation is done using the {\sc corsika} Extensive Air Shower Simulation code (\url{http://www-ik.fzk.de/corsika/}).

This paper is organized as follows: in Sec.\ \ref{sec:primary CR} we introduce the cosmic-ray spectrum and composition, and in Sec.\ \ref{sec:hadronic_ia} the hadronic interaction models. In Sec.\ \ref{sec:calculation_details} we describe the {\sc corsika} Monte Carlo calculation of lepton production in extensive air showers. In Sec.\ \ref{sec:results} we show the simulation results with comparisons with other calculations and direct experimental observations. First interaction models are benchmarked using muon observations, then the corresponding uncertainties on atmospheric neutrino spectra are determined. The effect on primary spectrum and composition is assessed as well.

\section{Primary Cosmic-Ray Spectrum}
\label{sec:primary CR}
\begin{figure*}[htbp]
	\centering
	\includegraphics[width=0.8\textwidth]{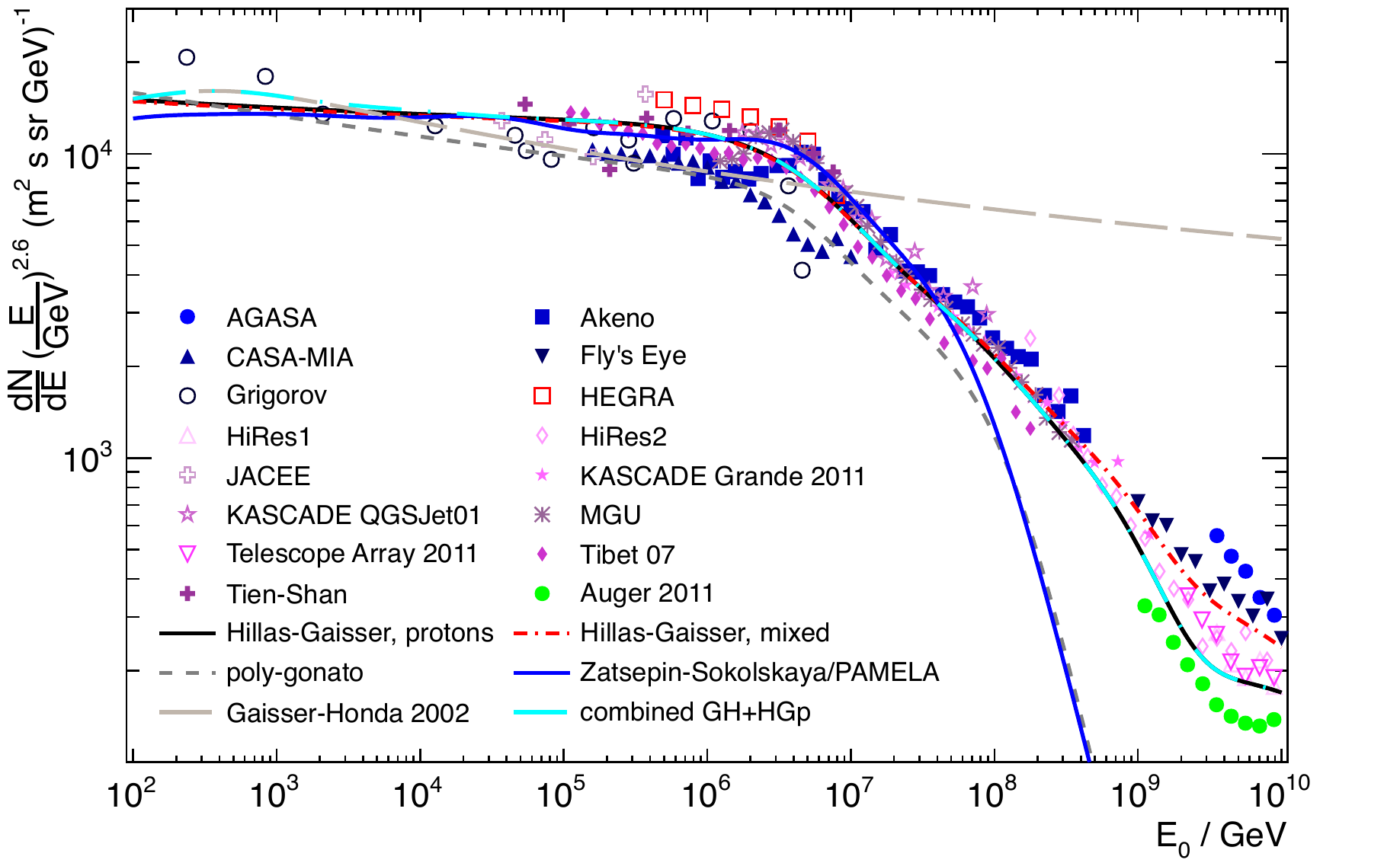}
	\caption{All particle spectrum as measured by ground based arrays. The data are from \cite{agasa,akeno,casa-mia,flys-eye,grigorov1,*grigorov2,hegra,hires,jacee1,*jacee2,kascade_grande_2011,kaskade,mgu1,*mgu2,ta_2011,tibet07,tien-shan}. The solid and dashed lines represent the power law models used as the parametrization of the primary cosmic-ray flux for this work. Data compilation after \cite{pdg_2011}.}
	\label{fig:primary_models}
\end{figure*}

The observed primary cosmic-ray flux covers a particle energy range from below $10^9$ eV up to several $10^{20}$ eV. In order to cover these 12 orders of magnitude in energy, a variety of different detection methods is used. Below $\sim$ 100 TeV particle energy, air-borne and satellite experiments such as AMS \cite{ams}, PAMELA \cite{pamela}, ATIC-2 \cite{atic-2}, CREAM \cite{cream1_2010,cream2_2010} and TRACER \cite{tracer}, directly measure particle energy and mass. Above $\sim$ 100 TeV, the cosmic-ray flux becomes too small and must be detected indirectly by large ground-based air-shower arrays (see Fig.\ \ref{fig:primary_models}). 
Different techniques such as the detection of secondary particles in an extensive air-shower or the measurement of air fluorescence are used. The fact that primary particles can only be observed indirectly by detecting portions of the induced air-shower makes it more difficult to measure the exact properties of the primary cosmic-ray spectrum. A number of very large surface arrays have
 been built to observe cosmic-rays up to about 10$^{20}$ eV. However, inconsistencies between the results of different experiments \cite{ta_2011}, presumably originating in the deviation of the simulation from the data \cite{auger_muon_deficit}, show that a better understanding of hadronic interactions at ultra high energies is needed.

The direct measurements provide the most unbiased results on cosmic-ray spectrum and composition, and great progress was made the past years. In particular, recent observations by  CREAM \cite{cream1_2010,cream2_2010} show an overall harder helium spectrum as compared to protons and, mainly, a flattening of their spectral index at about 230 GeV/nucleon. This result was confirmed by PAMELA \cite{pamela} and it is consistent with AMS results \cite{ams}. This flattening of the cosmic-ray spectrum could be an effect of particle propagation through the Galaxy or of acceleration in their sources \cite{cream_interpretation,malkov_2012}.

The indirect measurements have provided the observation of cosmic-ray spectrum at the knee region (which is at about $3\times 10^{15}$~eV, where it steepens from $\sim E^{-2.7}$ to $\sim E^{-3}$) and up to the highest energies. Recent measurements by KASCADE-Grande~\cite{kascade_grande_2011} are a first observational hint that the knee in heavy particles shifts towards high-energy, in general agreement with the notion of rigidity-dependence, although the result depends on the assumed hadronic interaction model. The highest energy cosmic-rays above the so-called ankle at $3\times 10^{18}$~eV are expected to be of extragalactic origin due to the high level of isotropy, which would not be present for Galactic sources \cite{auger_gc2007}.

In order to have a description of the expected atmospheric neutrino flux arising from cosmic-ray interactions in the Earth's atmosphere, a careful parametrization of the cosmic-ray composition is necessary. Some of the previous Monte-Carlo calculations reached neutrino energies up to 10 TeV e.g.\ \citet{bartol_2004} who uses a primary spectrum from \citet{agrawal_1995} or \citet{HKKM_2011} who use BESS \cite{BESS_TeV} and AMS \cite{ams} cosmic-ray data. Analytical calculations above the PeV region were performed by \citet{sinegovsky_icrc_2011}. In particular the latter uses the model of the cosmic-ray spectrum following \citet{zs_spectrum} (ZS), who assumed three classes of Galactic sources. The first source class is the explosion of Supernovae into the interstellar medium, the second class is motivated by the explosion of supermassive stars into the local super-bubble and the third class explains the flux of nuclei below 300 GeV by Nova explosions. The ZS model provides a smooth transition from the all-particle spectrum measured in the direct experiments to that measured with extensive air showers, and it is compatible with the all-particle spectrum by KASCADE  \citep{kaskade} and GAMMA \cite{gamma_composition}. All considered models with a (rigidity-dependent) knee are motivated by the fact that both acceleration and propagation in models involving collisionless diffusion in magnetized plasmas lead to the expectation of a rigidity-dependent cutoff for each individual component with a particle charge $Z$, $E_{cut,Z}\propto Z$~\citep{peters1961,ter_antonyan2000,ter_antonyan2001,stanev1993,ptuskin_1993}. This can explain the steepening of the spectrum around the knee and can be taken into account in the modeling of the cosmic-ray spectrum using a smoothed power law function as summarized in e.g.~\cite{ter_antonyan2000, Horandel_2003}. To effectively describe the all-particle spectrum of cosmic-rays, five different primary mass groups, namely H, He, CNO, Mg-Si and Fe, are usually used to obtain a realistic representation, see e.g.\ \citep{nikolskii1984,erlykin1987}. The individual spectra of the five components are summed up to get the all-particle spectrum. Recently, the PAMELA Collaboration has provided a new set of parameters for the proton and helium components of the first and third source class of the ZS model. These parameters are derived through a fit to their data \cite{pamela}. The agreement to the data is significantly improved, thus in the following we use these updated parameters and refer to the model as (ZS/PAMELA).

The poly-gonato model \cite{Horandel_2003,ter_antonyan2000} describes the individual mass spectra up to the knee region fairly well, nevertheless the relatively steep dependence above the knee is not in agreement with the all-particle spectrum observations above about 10$^{17}$ eV. Primary particles in this energy range contribute to the production of leptons in extensive air showers in the 100 TeV to PeV region. It is still disputed whether at this energy extragalactic cosmic-rays can be considered as valid source class, or if a second Galactic component contributes to the primary spectrum between the knee and the ankle. In \citet{hillas2006} it is suggested that the primary cosmic-ray spectrum is composed of three populations. The first population is associated to particles accelerated in Supernova Remnants with the knee indicating the cutoff. The second population (the so-called component B) accounts for the flux between the knee and the ankle and is associated with an unknown (or rather still debated) Galactic cosmic-ray population. The third population is assumed to be of extragalactic origin and becomes dominant above about 10$^{18}$ eV.

A recent ansatz including all three populations with up-to-date information on the composition of the spectrum is presented in~\cite{tom_primary_muoncharge}. In this model each population is represented by 5 mass groups with a simple power law and an exponential term representing the rigidity-cutoff. The spectrum for the first population is from the CREAM results extrapolated to a rigidity of about 4 PV to describe the knee, and the second population compensates for the all-particle disagreement between the knee and the ankle up to a rigidity of 30 PV. The extragalactic component is taken into account and two different composition scenarios are investigated. The first one testing an all-proton approach (HGp) and the second one using a mixed composition (HGm).

In this paper, we will pursue the following approach: to obtain a realistic representation of the composition we follow the common approach to use the five-component model. Since the observed hardening of the primary spectra is incorporated into the HGp/HGm models, we prefer to use these as the default high-energy model for this work. Both models are valid for energies above 10 TeV per particle. For a valid representation of the spectrum for energies below 10 TeV per nucleus, we extend the HG models to lower energies through a combination with the Gaisser-Honda 2002 (GH) \cite{gaisser_honda} spectrum. The transitions between the two spectra are calculated taking the the high He contribution as: 1500 GeV for H, 2650 GeV for He, 7280 GeV for CNO, 45.8 TeV for MgAlSi and 5050 GeV for Fe. In order to obtain a cross-over for the individual MgAlSi fits of the two models, the parameters of the GH model are shifted within the given error boundaries, resulting in a different normalization constant $K_{Mg-Si} = 34.2-6$ and a spectral index $\alpha_{Mg-Si} = 2.79 + 0.08$. The air shower simulation is run for each component individually and it produces relative abundances of the secondary particles (muons and neutrinos). The results can then be weighted with different models of primary cosmic-ray fluxes.

The resulting all-particle spectra predicted by the various models are superimposed on air shower data in Fig.\ \ref{fig:primary_models}.

\section{Hadronic interaction models}
\label{sec:hadronic_ia}
When a cosmic-ray proton or nucleus enters the Earth's atmosphere, it can be treated as a projectile colliding with an air nucleus target. An interaction with their hadronic constituents results in the production of mesons and baryons which spread away from the original projectile track with transverse momentum $p_\bot$. Unstable particles decay, while longer lived particles such as charged pions can undergo further interactions $\pi^\pm + Air \to$ anything, so representing a subsequent hadronic interaction with a different projectile-target configuration. The interplay of particle production, propagation and decay are the main ingredients of the atmospheric cascade. 

Air shower cascades are dominated by the soft component of the interaction, which represents the hadronic cascades with small transverse momentum with respect to the shower axis. Since no large momentum transfer is involved, the running coupling constant is too large for the application of ordinary perturbation theory \cite{dual_parton_model}. Due to the lack of a self-contained theory for the soft phase space region, it is usually attempted to describe the physics with phenomenological approaches, such as the Regge theory. 

The air shower simulation code {\sc corsika} \cite{corsika_report}] includes various models for low- and high-energy interactions. For this study only the high-energy interaction models are of interest, since the typical transition energy between high- and low-energy regime in {\sc corsika} occurs at the low energy boundary of this calculation at $80$ GeV. We have selected the models {\sc sibyll} 2.1 \cite{sibyll_1994}, {\sc qgsjet-01c} \cite{qgsjet01} with an additional heavy flavor (charmed) component and {\sc qgsjet-ii}-03 \cite{qgsjetII_3}, the successor of qgsjet-01. The restriction to these models is based on the acceptance in the air-shower community, the availability in {\sc corsika} and the computational time. While it would be interesting to test EPOS \cite{epos} it could not be done, since it demands approximately 60 times more calculation time than qgsjet-01.

The Quark-Gluon-String models with minijet production (qgsjet) are based on the phenomenological description of nuclear and hadronic collisions in Gribov's reggeon framework as multiple scattering processes \cite{qgsjet01,qgsjetII_3,qgsjetII_2,qgsjetII_1}. The individual scattering contributions, corresponding to microscopic parton cascades, are described in terms of the exchange of "soft" and "semihard" pomerons. The Glauber approach is used for hadron-nucleus interactions. The original qgsjet model as well as {\sc qgsjet-ii} were especially designed for cosmic-ray interactions, with the emphasis on the extrapolation to ultra-high-energy cascades. {\sc qgsjet-ii} has been extended by the treatment of non-linear effects concerning very high energies and small impact parameters, where a large number of elementary scattering processes occurs and the underlying partonic cascades strongly overlap and interact with each other. An implication of this approach for hadron-nucleus and nucleus-nucleus interactions is that the non-linear screening effects are stronger in nuclear case, breaking the superposition picture. Additionally, the parameters of {\sc qgsjet-ii} are tuned according to more recent accelerator data corresponding to the state of 2006. Regarding the execution performance, {\sc qgsjet-ii} is 20 times slower than qgsjet-01. 
The heavy flavor generation of {\sc qgsjet-01c} results in the production of charmed hadrons, so the description of the prompt component of the atmospheric lepton flux becomes possible. The model can handle charmed hadrons with the lowest mass, i.e. neutral and charged $D$($\overline{D}$) mesons and $\Lambda_c$($\overline{\Lambda_c}$) baryons \cite{qgsjet_charm,corsika_charm}. These particles are explicitly transferred to the propagation code, given the chance for interaction with air nuclei, which is rather uncertain due to the insufficiently known cross-sections. In {\sc corsika}, the decay is handled by \textsc{Pythia} 6.4 \cite{pythia} routines. Although, we have used this feature to explore the possibilities of the calculation of the prompt flux within a full Monte-Carlo approach, we would like to emphasize that for a detailed treatment of the charm contribution in air showers, one should refer to e.g. \cite{thunman_1996} or \cite{enberg_2008} and the references therein. 

The underlying physical model in {\sc sibyll} 2.1 \cite{sibyll_1994} is the Dual Parton Model \cite{dual_parton_model} for soft interactions with a mini-jet extension for the hard perturbative component \cite{dpm_with_jets}. Some features of the underlying code are borrowed from the Lund Algorithms contained in the \textsc{Pythia} 6 code. This model is explicitly optimized for air shower simulations, i.e.\ it implements extrapolation algorithms to ultra-high energies. Also, the program is efficiently designed to be called with a sequence of random collision energies and projectile-target configurations, in contrast to typical collider event generators, which expect to produce many events for the same configuration of incoming particles at the same center of mass energy.

The mechanism of fragmentation/hadronization is similarly treated in {\sc sibyll} and qgsjet. Both models describe the creation of new quarks and gluons in terms of one dimensional relativistic strings (color flux tubes), with one end attached to a valence (di-) quark from a projectile and a valence (di-) quark from the target, symbolizing the exchange of very soft gluons. The typical energy ("mass") density of such strings is in the order of $\kappa \approx 1 $ GeV/fm \cite{pythia}. When the distance between the partons exceeds a certain critical value, the string breaks, creating a $q\overline{q}$ pair. The algorithm assigns each of these quarks to the open string ends, retaining the color confinement. Using this approach it is possible to explain the creation of new hadrons until the total energy stored in the original string has been dissipated as mass and forward momentum.

Regarding the description of heavy nuclei collisions, {\sc sibyll} employs the Glauber approach for the description of hadron-nucleus interactions and implements a semi-superposition picture \cite{semi_superposition} for nucleus-nucleus interactions. The qgsjet models employ the Glauber-Gribov approach for nucleus-nucleus interactions, taking into account inelastic screening effects. In this case the nucleus-nucleus cross sections are noticeably reduced compared to the pure Glauber case. However, the calculation method of this work is based on averages of shower observables, so that the result will likely converge towards the super-position picture \cite{gaisser_book,qgsjet01}.

\section{Overview of calculation}
\label{sec:calculation_details}
The central element of the calculation is the Extensive-Air-Shower (EAS) simulator {\sc corsika} version 6.990. We select a spherical detector to observe the theoretical flux without restricting the calculation to a certain type of detector. The simulation proceeds by injecting primary nuclei (H, He, C, Si and Fe) at the top of the atmosphere, such that propagation, hadronic interactions and particle decay are handled within the unmodified {\sc corsika} code. The temperature/density profile of the atmosphere is modeled by 5 exponential functions, with parameters fitted to the US Standard Atmosphere \cite{us_std_atmosphere}. This static atmospheric model is widely used in neutrino flux calculations as a representation for the global atmosphere see \cite{HKKM_2004,HKKM_2007,HKKM_2011}, or \cite{wentz_2003} for an overview. Throughout the simulation, the Earth's curvature is taken into account as described in \cite{ corsika_curved}.

At lepton energies below hundreds of TeV the main sources of muons and neutrinos are decays of charged pions and various types of kaons ($K^\pm$, $K^0_S$, $K^0_L$). In the latter case two- and three-body decay modes are included. Using the hadronic generation counter in {\sc corsika}, it is only possible to identify leptons having a pion as mother particle or not, thus chained decays  such as $K \to \pi^\pm \to \mu + \nu_\mu$ are identified as pure pion decays.

An interesting production channel of prompt muons via the decay of the $\eta$-mesons is claimed by the authors of \cite{illana_eta_2009,illana_eta_2010} to be the dominant source (di-)muons for energies in the PeV range. From our own calculation of the energy dependent z-factors using {\sc sibyll}-2.1 we can estimate that this channel has importance and the product $Z_{p \eta} \times BR_{\eta \to \gamma \mu^+ \mu^-}$ is indeed in the same order of magnitude as the corresponding factors for the charm production via $D$-mesons obtained from PYTHIA in \cite{thunman_1996}. However, the corresponding decay channel $\eta \to \gamma \mu^+ \mu^-$ is not available in the present {\sc corsika} version, thus we can not verify it's importance within this MC calculation.

Also, due to constraints on the computational time the simulation of the electro-magnetic component of EAS has not been activated, thus the pair production of muons via $\gamma \to \mu^+ + \mu^-$ is not included here. The influence of this channel to the total muon flux was estimated by the authors of \cite{illana_eta_2010}, to be an order of magnitude lower compared to their prediction of the muon flux from $\eta$-mesons.

\subsection{Calculation scheme and normalization}
When the secondary particles reach the sea level, they are binned according to their energy $E$, type $p$, mother particle $m$ ($K$, $\pi$ or charm), the high-energy interaction model $\mathcal{M}$, the discrete energy of the primary nucleus $E_0$, the charge of the primary nucleus $\mathcal{Z}$, the zenith angle of the primary nucleus $\theta$ and the atmosphere $\mathcal{A}$. This results in a database of differential inclusive energy spectra $(dN/dE)$, called yields $\mathcal{Y}_{m \to p}( E_{p}, \mathcal{M}, \mathcal{A}, \theta, \mathcal{Z}, E_0)$.

The resulting flux of leptons of type $p$ at the surface can be calculated as the discrete convolution
\begin{eqnarray}
\label{eq:convol} \Phi_{p}(E_{p}, \theta, \mathcal{M}, \Phi_{\mathcal{C}}, \mathcal{A}) &=& \sum_{m_k}\sum_{\mathcal{Z}_j}\sum_{E_{0,i}} w(E_{0,i},\Phi_{\mathcal{C}}(\mathcal{Z}_j)) \\
\nonumber &\times&  \mathcal{Y}_{m_k \to p} (E_{p}, \mathcal{M}, \mathcal{A}, \theta, E_{0,i}, \mathcal{Z}_j) ~,
\end{eqnarray}
with the weight function 
\begin{equation}\label{eq:weight}
w(E_0,\Phi_{\mathcal{C}}(\mathcal{Z}_j)) = s_A \cdot s_N(E_0,\Phi_{\mathcal{C}}(\mathcal{Z}_j)) ~ .
\end{equation}
The surface scaling factor $s_A$ compensates for the area normalization due to different reference shells for flux of cosmic-rays at the top of the atmosphere relative to the area of the virtual detector at the surface \citep{wentz_2003}:
\begin{equation}
s_A = \frac{r_E + h_{atm}}{r_E} \approx 1.018~,
\end{equation}
where $r_E$ is the Earth's radius and $h_{atm}$ is the atmosphere's height.
The second factor $s_N(E_{0,i},\mathcal{Z}, \theta)$ compensates for the number of simulated showers $N$ in the $i$-th primary energy bin $\Delta E_{0,i}$, with respect to the physical flux in this energy bin according to some theoretical primary flux model $\Phi_{\mathcal{C}}$:
\begin{equation}
s_N(E_{0,i},\Phi_{\mathcal{C}}(\mathcal{Z})) = \frac{1}{N(E_{0,i})}  \int_{\Delta E_0,i} dE'_0~\Phi_{\mathcal{C}}(E'_0,\mathcal{Z}).
\end{equation}
The factor is calculated for each primary energy bin $i$ and nucleus $\mathcal{Z}$. In this work $N(E_0 = 100\ \mathrm{GeV}) = 100 000$, extending with a power law behavior ($\sim E^{-\gamma}$) up to $N(E_0 = 100\ \mathrm{EeV}) = 40 000$. This corresponds to a  differential spectral index of $\gamma = 1.05$. This approach allows us to efficiently re-weight the simulated dataset according to any primary model without the recomputation of the air shower database.

\subsection{Approximations}
For secondaries above 80 GeV the 3D-effect, the east-west effect and the up-down asymmetry have a negligible contribution \citep{HKKM_2004,HKKM_2007,bartol_2004}. Also, for primaries above hundreds of GeV the influence of the geomagnetic cutoff and the Earth's magnetic field is below the simulation accuracy. Therefore, we make following approximations:
\begin{enumerate}
  \item the flux of primary cosmic-rays at the top of the atmosphere is isotropic,
  \item the error due to joint usage of the US Standard atmosphere is smaller than other systematic uncertainties,
  \item the effect of Earth propagation is neglected, thus the flux of neutrinos is assumed to up-down symmetric up to $\sim$ PeV energies (see e.g.\ \cite{icecube40_diffuse,neutrino_earth_prop} for the estimation of this effect on the event rates of neutrino telescopes)
  \item the flux is azimuth-symmetric,
  \item neutrino oscillations have no significant effect,
  \item the lateral distribution of the EAS is not taken into account and particles are considered to travel on a line trajectory, even if the full tree-dimensional cascade is simulated.
\end{enumerate}
Since in {\sc corsika} the shower has a fully three-dimensional shape, the last approximation is motivated by the disregard of the lateral distribution. The total number of simulated showers is $787 161 600$.

\subsection{A semi-analytical approximation}
\label{ssec:semianalytical}
A comparison of the Monte-Carlo to the semi-analytical solution of the cascade equations requires a correct treatment of the energy behavior of the interaction models and a non-powerlaw primary flux to be represented by the result. The concept, introduced in \cite{thunman_1996}, which allows to assess both points is the energy-dependent z-factor
\begin{equation}
Z_{kh}(E) = \int_E^\infty ~ \rm{d}E' \frac{\phi_k(E',X,\theta)}{\phi_k(E,X,\theta)} \frac{\lambda_k(E)}{\lambda_k(E')}\frac{dn(kA \to hY;E',E)}{dE},
\end{equation} 
where $k$ stands for the particle entering the hadronic interaction, $h$ the inclusively produced hadron, $\phi_k$ the flux of primary cosmic-ray nucleons evaluated at the energy of the hadron, $\lambda_k$ the interaction length in air and the last factor is the inclusive spectrum if hadrons $h$ produced in interactions of particles $k$ of energy $E'$ with air.
Due to $\lambda_k(E) \propto 1/\sigma_{kA}(E)$ in can be written as
\begin{equation}
Z_{kh}(E_l) = \int_E^\infty ~ \rm{d}E' \frac{\phi_k(E',X,\theta)}{\phi_k(E,X,\theta)} \frac{\sigma_{kA}(E')}{\sigma_{kA}(E)}\frac{dn(kA \to hY;E',E)}{dE}.
\end{equation}
To reduce the differences respective the {\sc corsika} calculation the hadron-Air, cross sections have been extracted from the interaction models' code. 

The lepton flux at the surface can be calculated using the asymptotic solutions of the cascade equation, interpolated between the low and high-energy solutions separated in a decay and an interaction dominated regime, respectively \cite{gaisser_book,Lipari:1993hd,thunman_1996}. The formula for $l=\nu_\mu, \nu_e, \mu$ is
\begin{equation}
\phi_l(E) = \frac{\phi_N(E)}{1 - Z_{NN}}\sum_{\pi,K,K^0_L} \frac{Z_{NM} Z_{M \to l, \gamma + 1}}{1 + A_M E \cos\theta^*/\varepsilon_M},
\end{equation}
with
\begin{equation}
A_M = \frac{Z_{M \to l, \gamma + 1}}{Z_{M \to l, \gamma + 2}}\frac{1-\Lambda_N/\Lambda_M}{\ln(\Lambda_M/\Lambda_N)}
\end{equation}
$M$ is the semileptonically decayed meson, $Z_{Ni}$ the energy dependent z-factors for proton or neutron interactions, $\Lambda_i$ the nucleon/meson attenuation lengths, and $Z_{M \to l, \gamma}$ the decay z-factors defined as in \cite{thunman_1996}, including the decay kinematics and the branching ratio. 
Only ratios of the hadron attenuation lengths $\Lambda_i$ are involved. Therefore, it is possible to replace their ratios with
\begin{equation}
\frac{\Lambda_i}{\Lambda_N} = \frac{\sigma_{NA}(E) (1 - Z_{NN})}{\sigma_{i A}(E) (1 - Z_{ii})},
\end{equation} 
since the interaction length $\lambda_i \propto 1/\sigma_{i A}$, and $i = \pi,K$. The cross-sections are taken from the interaction models. 

The kaon channel has a more difficult context, since $K^+$ and $K^-$ are not in the same isospin group and threrfore their production rate is not symmetric. In addition the evolution equations of the different kaon species, including $K^0$, are coupled \cite{Lipari:1993hd}. For the purpose of this calculation it should be adequate to assume that all relevant kaons are produced in the first interaction of the cosmic-ray nucleon. The regeneration and the strangeness changing contribution of the different kaon species during the cascade evolution are therefore neglected. This results in follwing z-factor relations: 
\begin{eqnarray}
\nonumber &Z_{NN}& = Z_{pp} + Z_{pn} \\
\nonumber &Z_{N\pi}& = Z_{p\pi^+} + Z_{p\pi^-} \\
\nonumber &Z_{pK}& = \frac{p}{\phi_N} (Z_{pK^+} + Z_{pK^-}) \\
\nonumber &Z_{nK}& = \frac{n}{\phi_N} (Z_{nK^+} + Z_{nK^-}) \\
\nonumber &Z_{NK}& = Z_{pK} + Z_{nK} \\
\nonumber &Z_{\pi\pi}& = Z_{\pi^+\pi^+} + Z_{\pi^+\pi^-} \\
\nonumber &Z_{KK}& = Z_{K^+K^+} + Z_{K^+K^-},
\end{eqnarray}
where $p$ and $n$ are the fraction of protons and neutrons of the cosmic-ray flux, respectively. The production of $\overline{K^0_L}$ is neglected. The decay z-factors are interpolated from the table provided in \cite{thunman_1996}.

The zenith angle is corrected for the curvature of the atmosphere in the $\cos{\theta}^*$ representation. The main idea is contained in \cite{Lipari:1993hd}. Here, we use a parametrization extracted from a {\sc corsika} Monte-Carlo \cite{dima_corsika}.

\section{Results}
\label{sec:results}
The structure of this chapter is as follows: in the first part the particle charge and flavor ratios are calculated, reflecting some physics assumptions of the employed interaction models and features of the air shower code. These values are mainly sensitive to the pion to kaon ratio in the air shower. The differences between the interaction models are dominant compared to the variation due to the primary cosmic-ray flux. Therefore, the calculation of the conventional ratios has been carried out, using the combined Gaisser-Honda and Hillas-Gaisser model with protons only in the extragalactic component (cHGp). If the zenith range is specified as vertical, then it is restricted to $\cos(\theta) < 0.25$ and  $\cos(\theta) > 0.75$ for horizontal, respectively. Generally, neutrino fluxes are given as averages over all zenith angles and muon fluxes for the vertical direction only.

In the second part, we present the calculated conventional fluxes of atmospheric muons, muon neutrinos and electron neutrinos. Here, we again use the cHGp model to parametrize the primary cosmic-ray flux and composition.

In the last part, we study the influence of the knee of cosmic-rays by variation of the primary CR flux model for a given interaction model. Additionally, we show the prompt prediction based on the implementation of {\sc qgsjet-01c} for different primary flux models.
 
\subsection{Interaction model performance}
\label{ssec:iam_performance}
The ratios of neutrino flavor and of muon charge represent the interaction model performance regarding the inclusive spectra of pions and kaons in the atmospheric cascade, where kinematical propagation effects, a realistic atmosphere and air composition are taken into account. There is also sensitivity to the primary composition, since a higher contribution of neutrons in the all-nucleon flux lowers the fraction of positively charged particles in the shower. The statistical uncertainty of the Monte-Carlo method limits the reasonable energy range to $< 100$ TeV. The errors are purely statistical.

\subsubsection{Lepton ratios at the surface}
\begin{figure*}[htbp]
	\centering
	\includegraphics[width=0.43\textwidth]{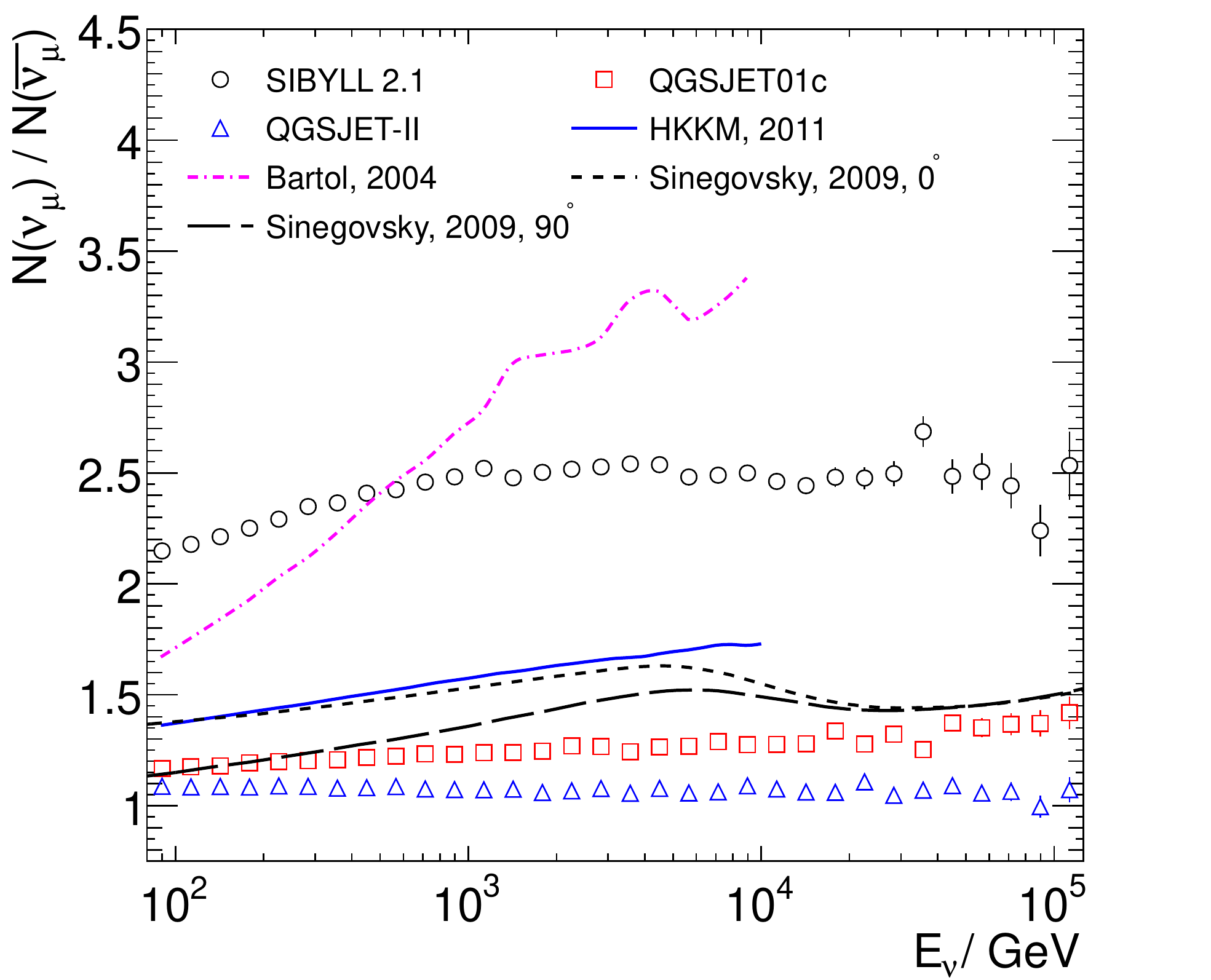}
	\includegraphics[width=0.43\textwidth]{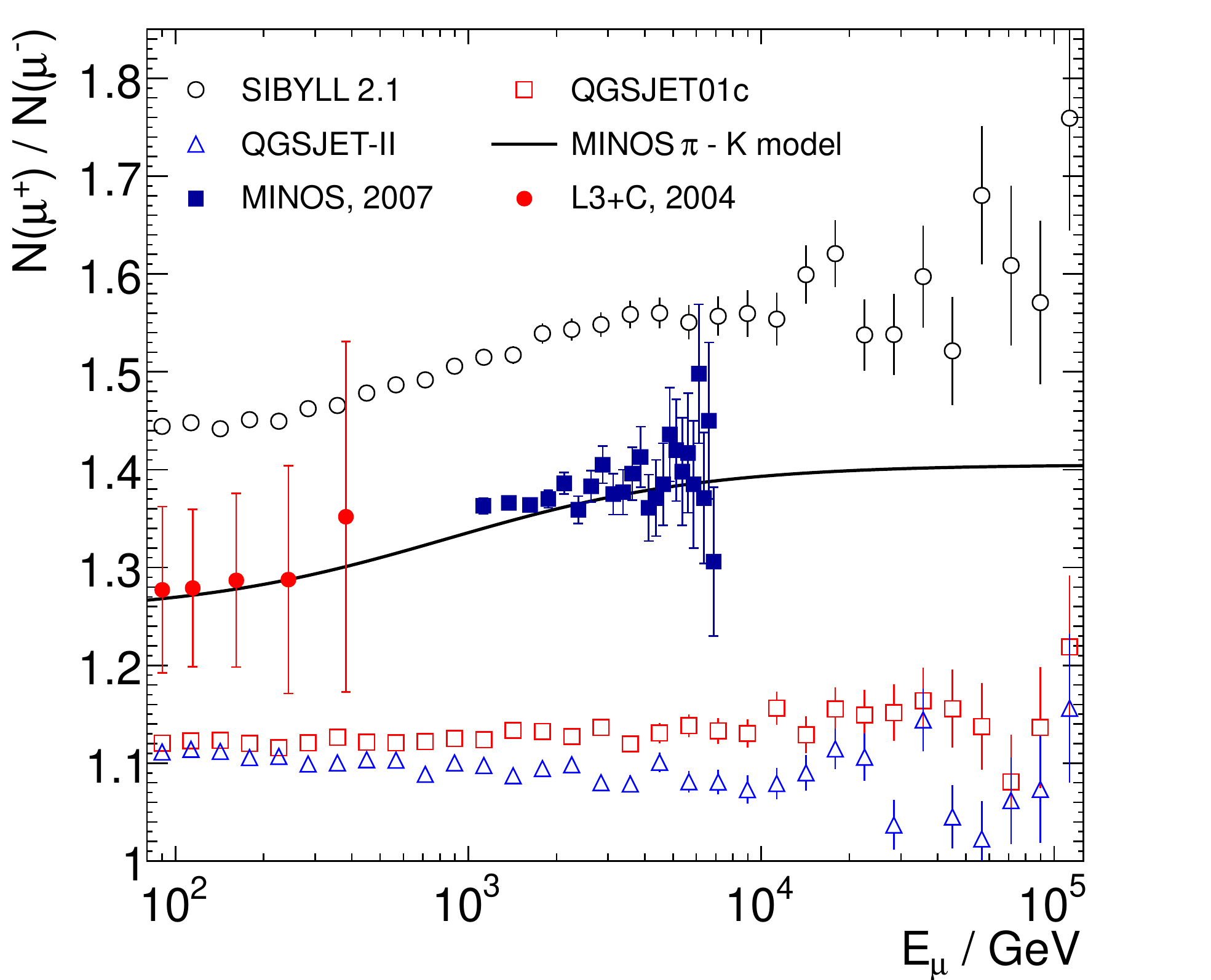} 
	\includegraphics[width=0.43\textwidth]{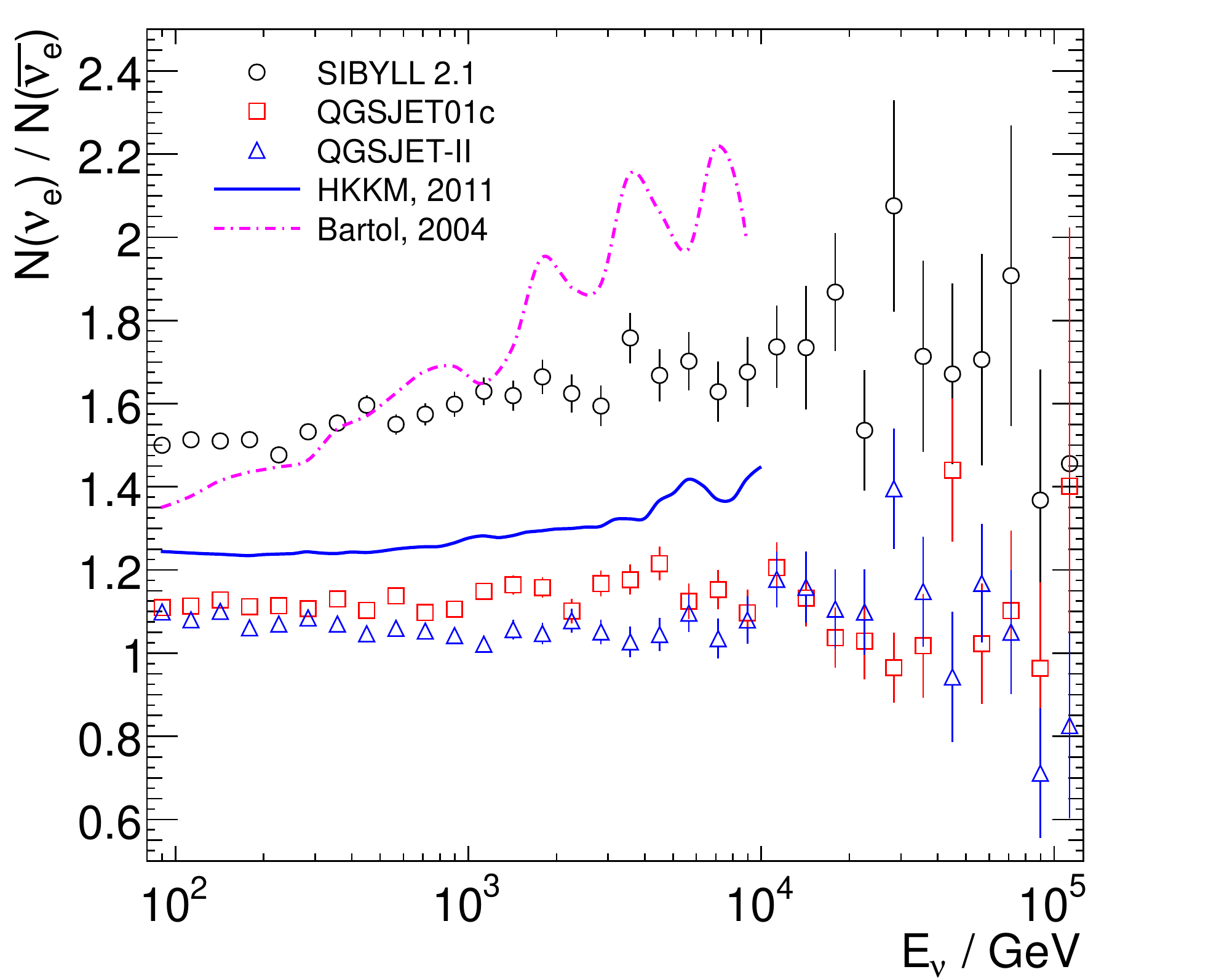}
	\includegraphics[width=0.43\textwidth]{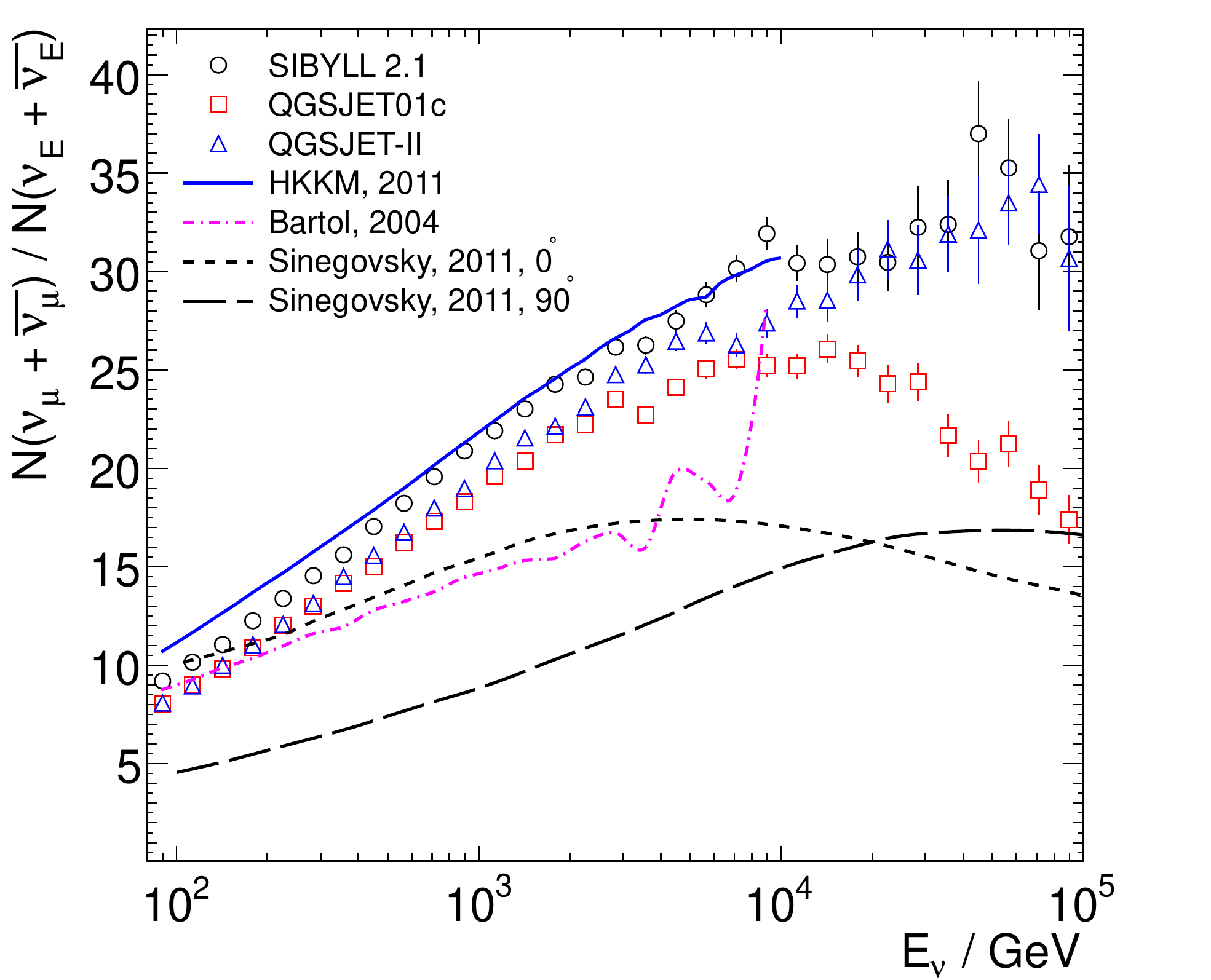}
	\includegraphics[width=0.43\textwidth]{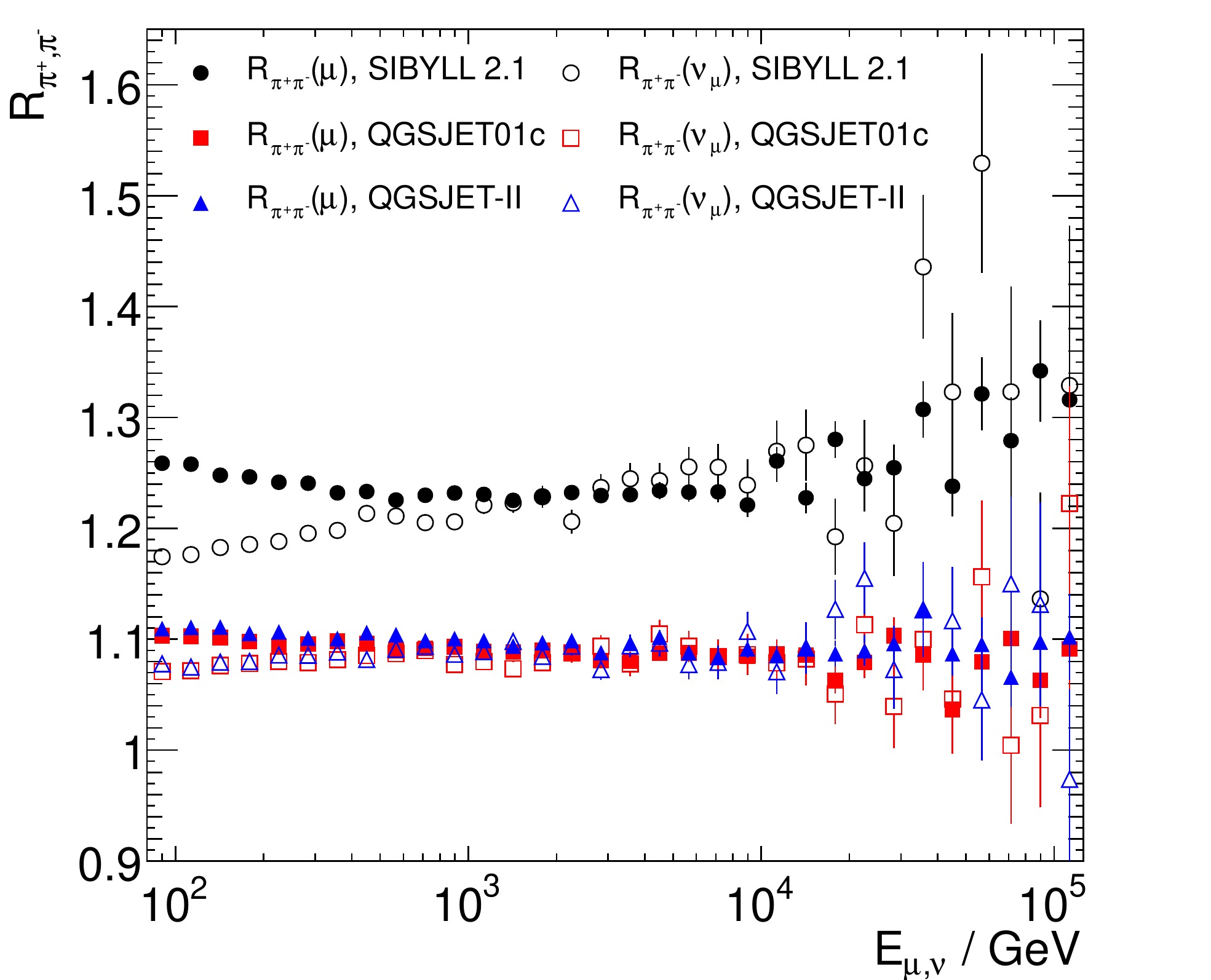}
	\includegraphics[width=0.43\textwidth]{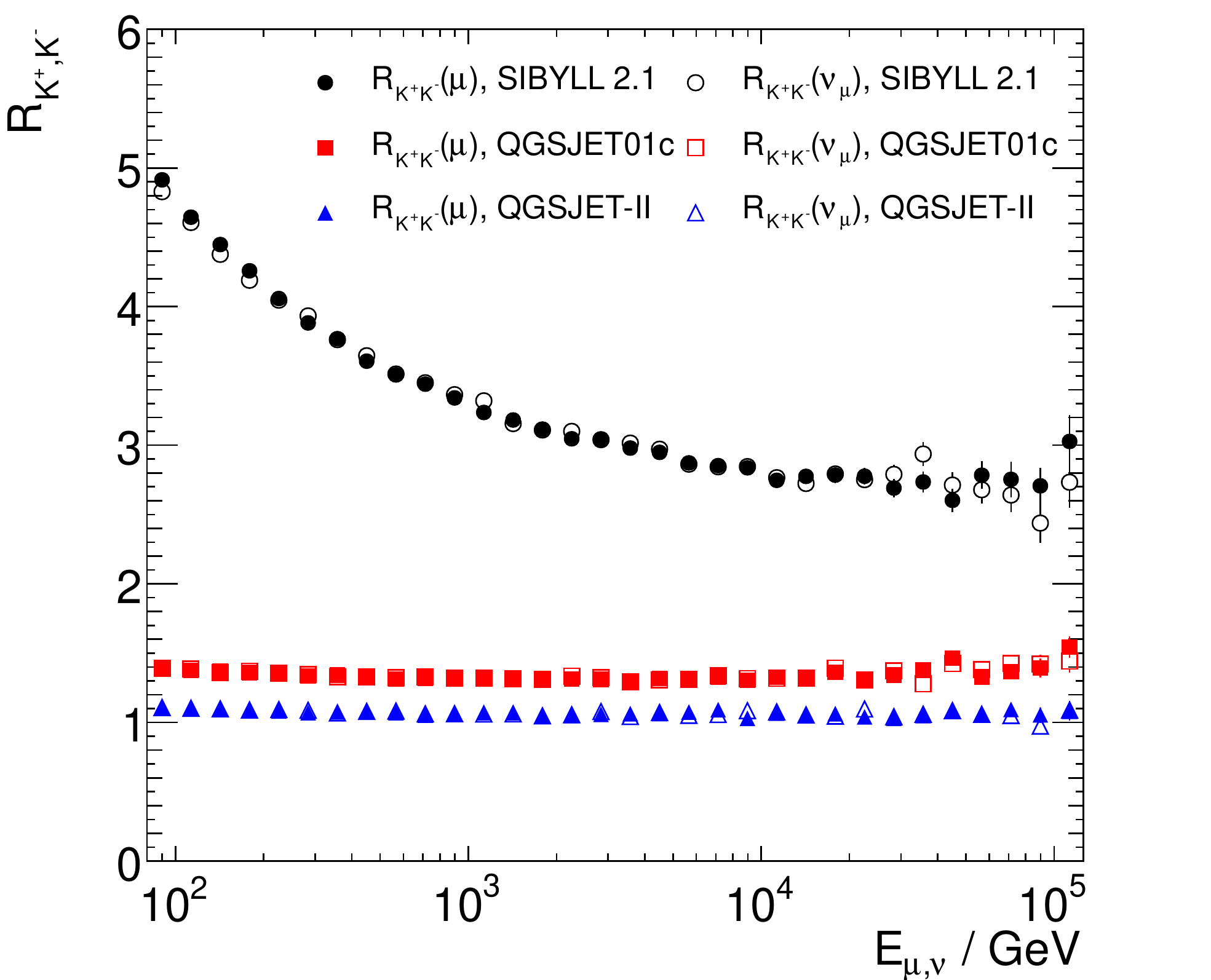}
	\caption{{\it Top left and middle panels}: All angle averaged ratios of neutrinos, compared to calculations \cite{bartol_2004}, \cite{HKKM_2011}, \cite{sineg_icrc_2009} and \cite{sinegovsky_icrc_2011}. {\it Top right panel}: Muon charge ratio for vertical muons, compared to data by MINOS \cite{minos_mucharge} and L3+C \cite{L3_C_muons}. {\it Bottom panels}: Charge ratio of pions (\textit{left}) and kaons (\textit{right}) derived from surface muons (full markers) and muon neutrinos (hollow markers), respectively.}
	\label{fig:lepton_ratios}
\end{figure*}

Figure \ref{fig:lepton_ratios} shows ratios of neutrino flavor and of muon charge as they would be observed by a surface detector.
In the upper left panel, the muon neutrino to antineutrino ratios are drawn together with some reference calculations. HKKM 2011 \cite{HKKM_2011}, which uses a modified version of DMPJET-III \cite{dpmjetIII} for the high-energy part, performs close to the cascade equation approach by \citet{sineg_icrc_2009}, employing  the parametrization of nuclear interactions by Kimel-Mokhov and the ZS primary flux model. The Bartol 2004 \cite{bartol_2004} calculation reflects the properties of the interaction model TARGET \cite{target_2.1}, which explicitly includes the associated kaon production channel $p + \rm{air} \to \Lambda + K^+ + \rm{anything}$ \cite{target_2.1,sibyll_1994,agrawal_1995}. The similarities between {\sc sibyll} and TARGET become apparent due to the higher kaon charge ratio and thus the increased $\nu_\mu$ over $\overline{\nu}_\mu$ flux, due to the similarly treated associated kaon production. Both qgsjet models fall below all other expectations because the charge ratios (bottom panels of Fig.\ \ref{fig:lepton_ratios}) of pions and kaons in the air shower are constant over a wide energy range, showing that no associated kaon production is included.

In analogy to the $\nu_\mu/\overline{\nu}_\mu$-ratio, the muon charge ratio for vertical zenith directions is shown in the top right panel of Fig.\ \ref{fig:lepton_ratios}. As expected from the muon neutrino ratios, the muon charge ratio from {\sc sibyll} suffers from the overestimation of $K^+$ or of the pion charge ratio. However, the curve reproduces the shape of the atmospheric muon charge ratio, which is cruicially influenced by the hierarchy of the involved meson lifetimes or their critical energies, respectively. It is therefore presumably a matter of scale between the inclusive pion and kaon spectra in the $p~N$-/$n~N$-interaction. Due to the flat meson charge ratios and also the low kaon charge ratio, both qgsjet models do not describe the shape and normalization according to L3+C and MINOS data correctly.

The middle left panel shows the $\nu_e/\overline{\nu}_e$-ratio. Again, the difference between the interaction models which have an increased $K^+$ production can be observed due to the production channel $p \to \Lambda K^+$ and the three-body decay $K^{\pm} \to \pi^0 e^\pm \nu_e(\overline{\nu}_e)$. Within {\sc sibyll} this process has a higher contribution (see bottom panels).

The neutrino flavor ratio in the middle right panel is sensitive to the occurrence of the 3-body kaon decay process $K_{3e\nu}$ \cite{Honda_1D_1995}, which is the dominant source of conventional electron neutrinos in this energy range. Our calculation is close to HKKM 2011 for all three interaction models. From \cite{sinegovsky_icrc_2011} the {\sc sibyll} + GH calculation is used. The falling ratio for {\sc qgsjet-01} at $E_\nu > 10$ TeV is from the additional $\nu_e$ flux, coming from decays of charmed hadrons.

The bottom panels show the charge ratio of the mother mesons of muons and neutrinos at the surface. In {\sc corsika}, secondary neutrinos are treated as final state particles without further interactions. The difference between the pion charge ratio for muon and neutrino energies below 1 TeV only occurs for horizontal zenith angles and thus it can be explained through the occurrence of chained decays of $K^+ \to \pi^+ +~\mathrm{anything} \to \mu^+ + \nu_\mu$ and muon decay $\mu^\pm \to e^\pm +\nu_e(\overline{\nu}_e) + \nu_\mu(\overline{\nu}_\mu)$. In the right panel, the kaon charge ratio is nearly constant for all energies and the qgsjet models (no associated production). In {\sc sibyll} the enhancement due to associated kaon production is apparent as stated above. This behavior suggests that the $\Lambda$ cross-section is too large or their spectrum too hard. This effect enhances the total kaon multiplicity in the cascade, leading to a higher neutrino flux at the surface. This behavior is reflected in the behavior of corresponding z-factors.

\subsubsection{Mesonic origin of leptons}
\begin{figure*}[htbp]
	\centering
	\includegraphics[width=0.43\textwidth]{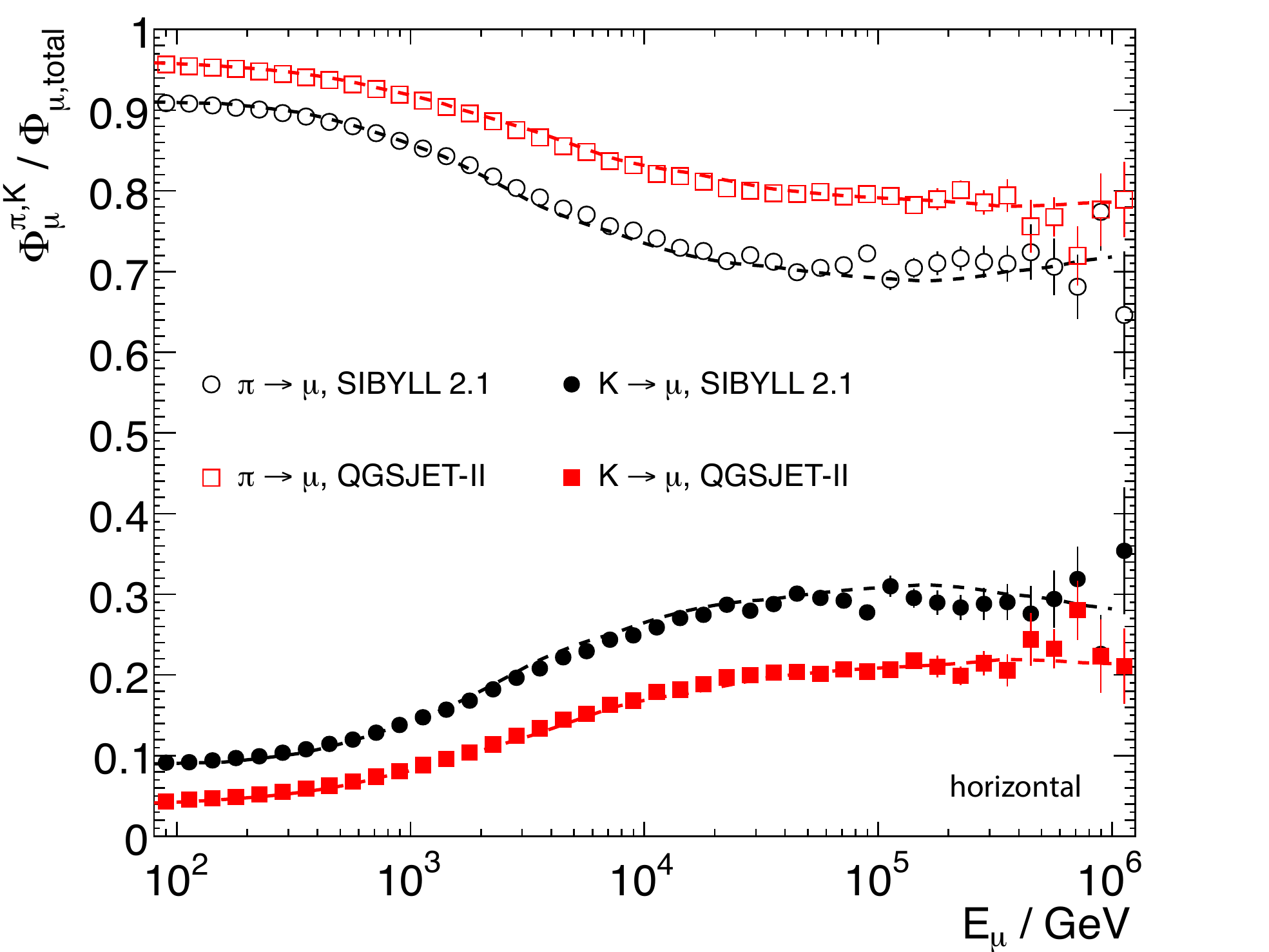}
	\includegraphics[width=0.43\textwidth]{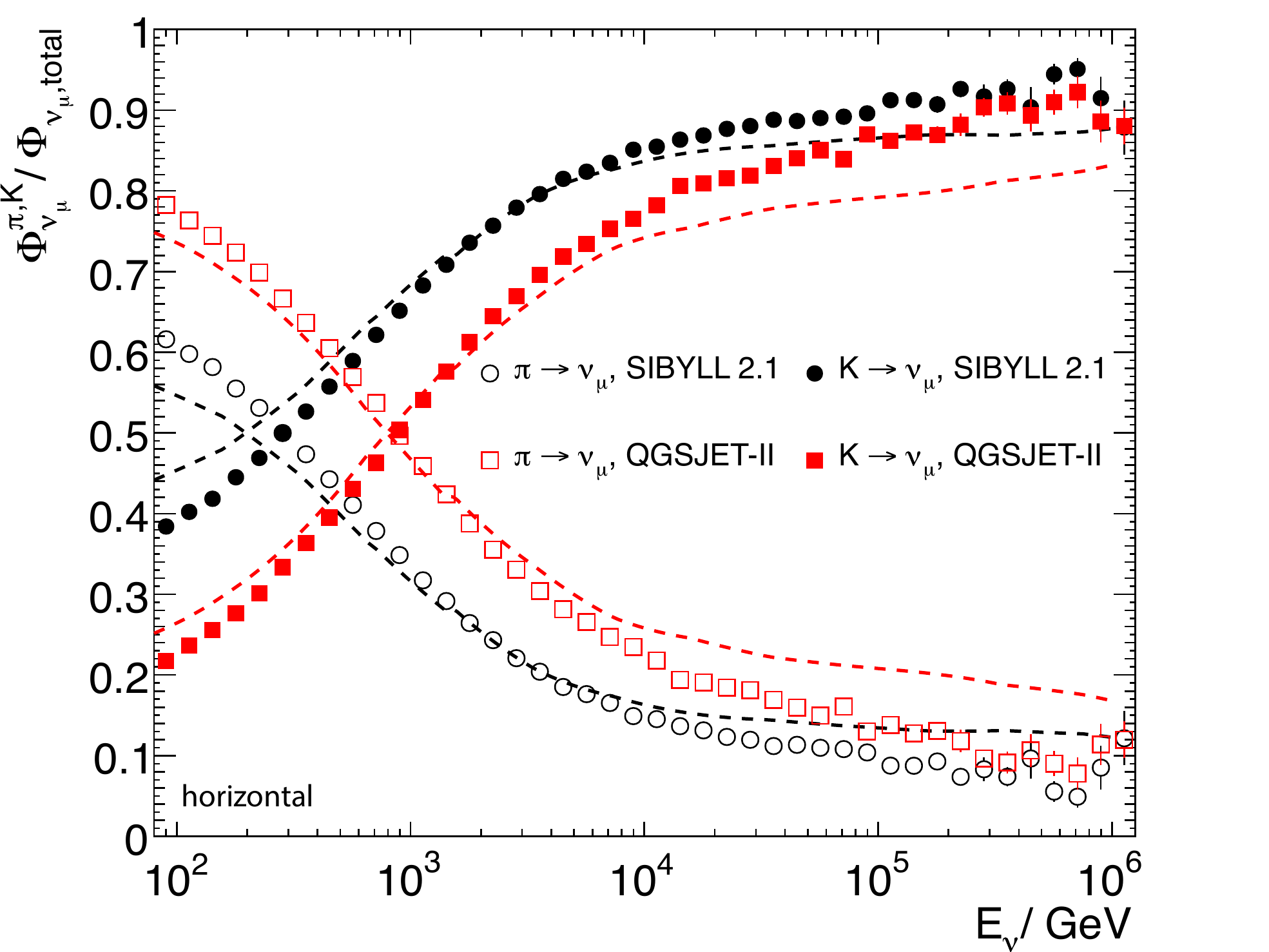}
	\includegraphics[width=0.43\textwidth]{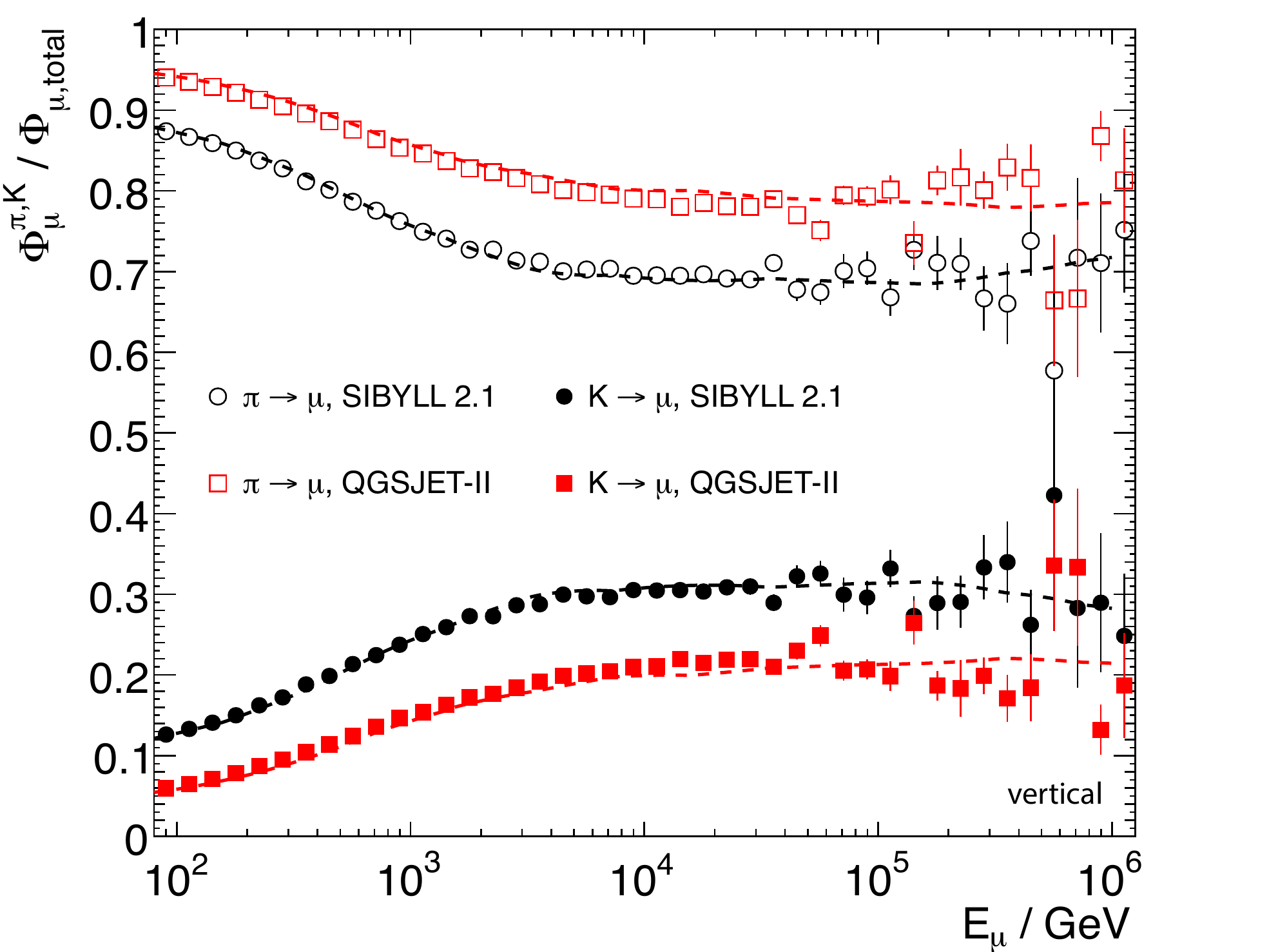}
	\includegraphics[width=0.43\textwidth]{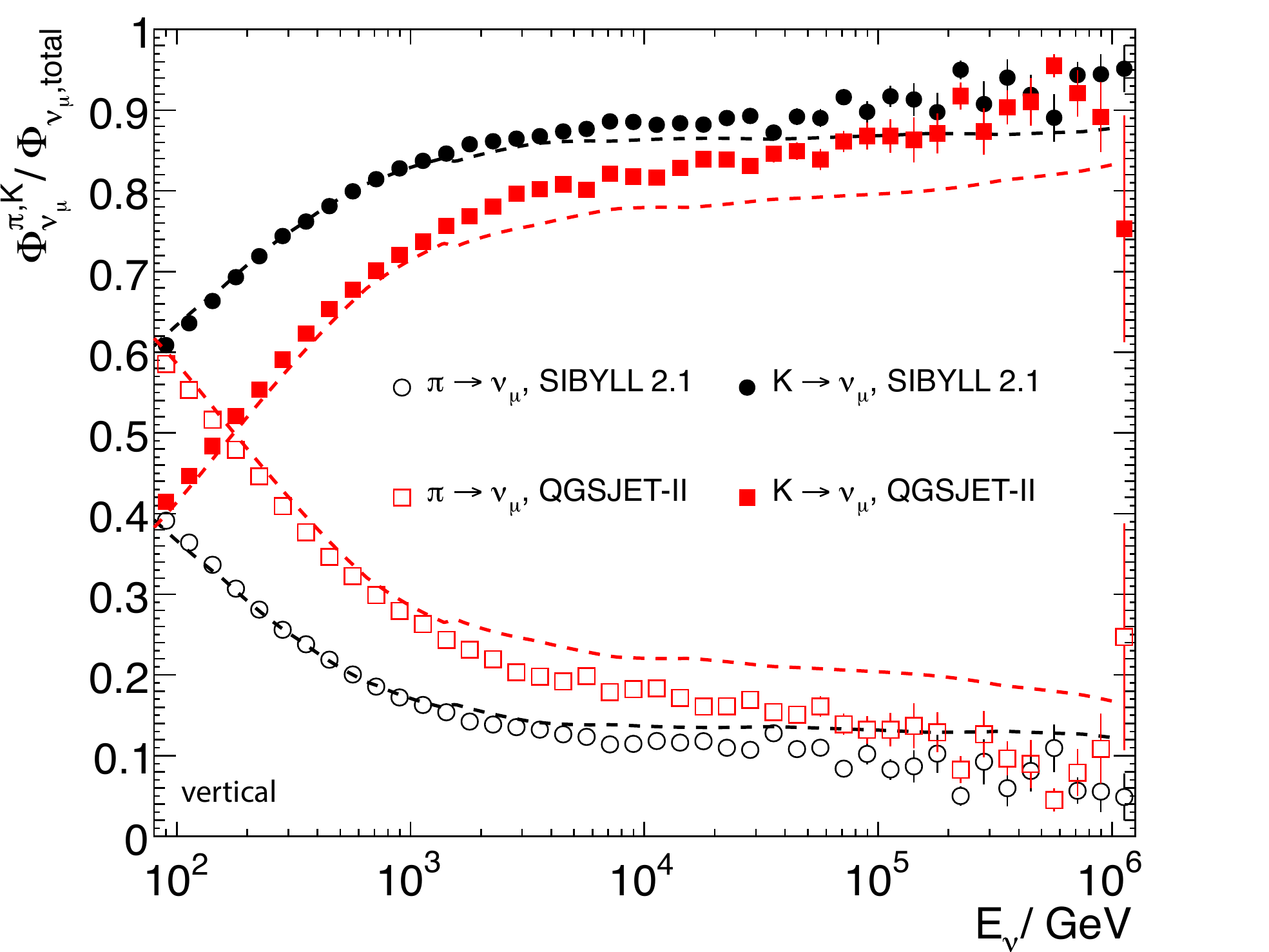} 
	\includegraphics[width=0.43\textwidth]{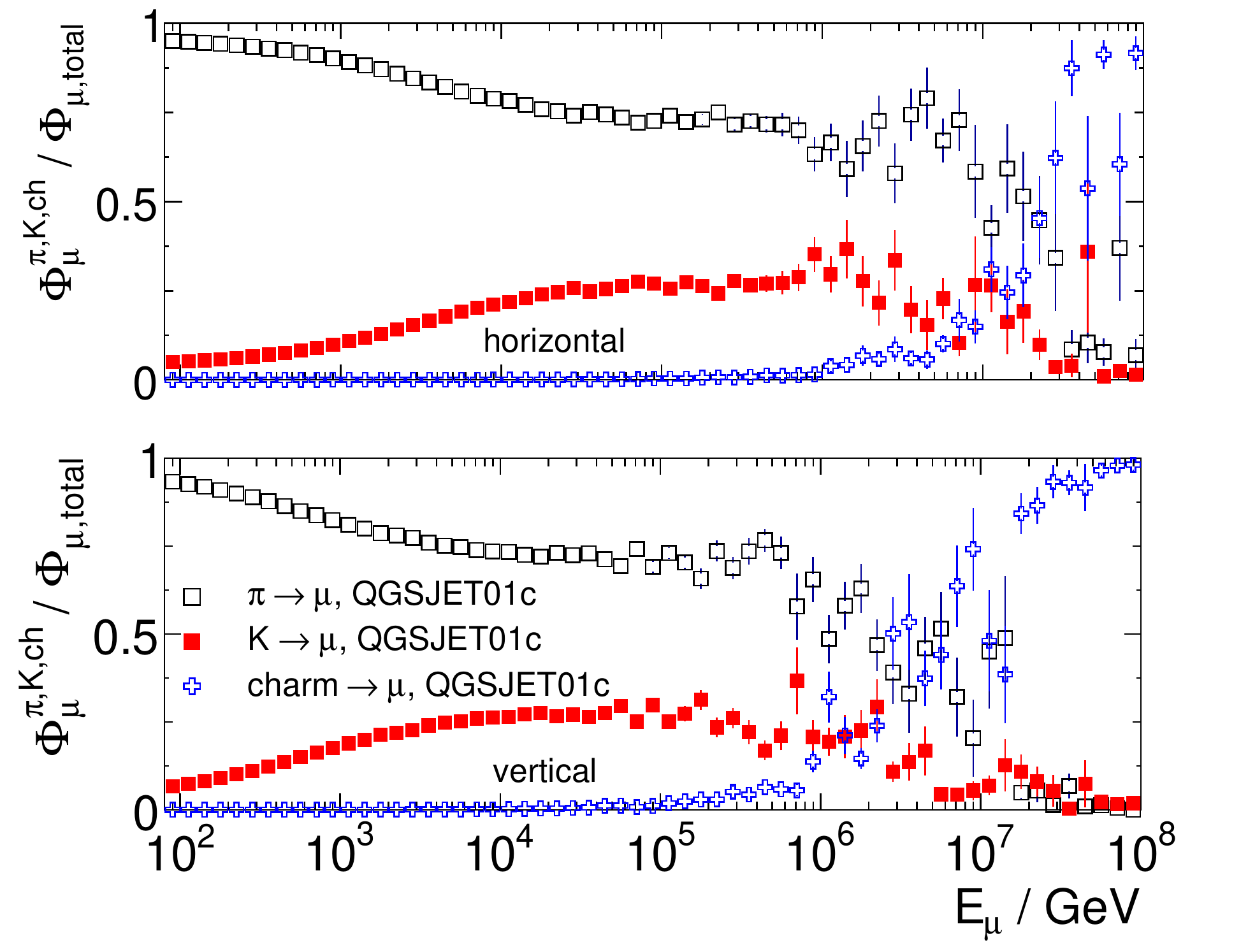}   
	\includegraphics[width=0.43\textwidth]{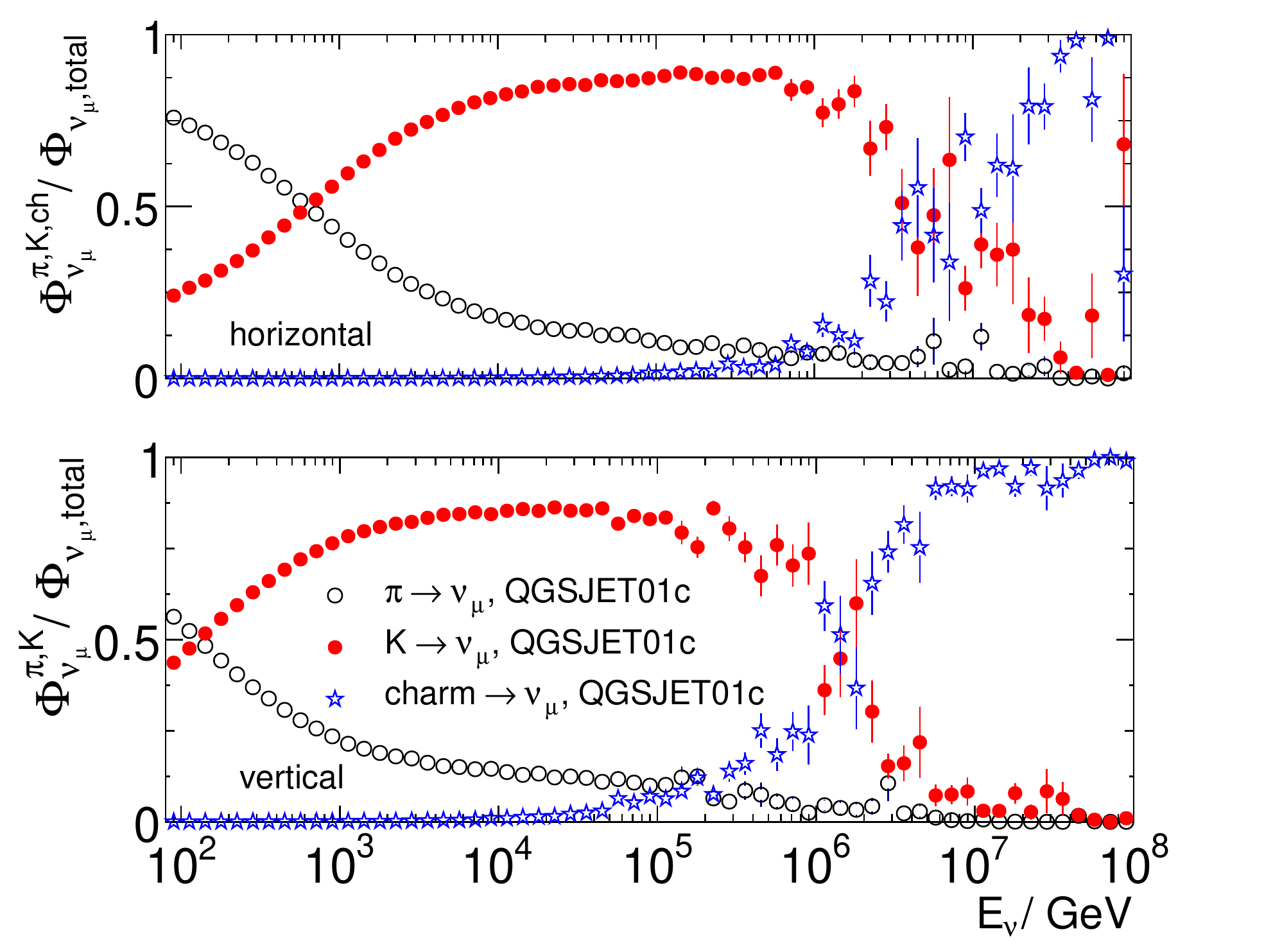}   
	\caption{Fractions of pions and kaons to the total flux for vertical and horizontal muons (\textit{left hand side}) and neutrinos (\textit{right hand side}), respectively. The dashed curves are obtained through the semi-analytical approximation described in section \ref{ssec:semianalytical}. The four bottom panels separately show the corresponding figures obtained with qgsjet01c including charmed hadrons.}
	\label{fig:k-pi-fractions}
\end{figure*}

In Fig.\ \ref{fig:k-pi-fractions} the fractions of muons and neutrinos are shown with respect to their mother particle of arbitrary charge. As expected, {\sc sibyll} has a higher contribution of kaons at all energies. 
For energies above $E_\nu > 100$ TeV the source of conventional neutrinos are nearly exclusively kaons. The cascade equation (CE) solutions satisfactory describe the fractional contribution of $\pi$ and $K$ to the flux of atmospheric muons. The same quatities for neutrinos differ slightly for higher energies. This can originate from additional energy loss mechanisms in the Monte-Carlo, or because source terms for secondary hadron interactions are not included in the CE.

The four bottom panels show qgsjet01c predictions including the prompt component. For the vertical incident direction the conventional flux is suppressed by the steeper atmospheric pressure gradient, such that longer lived mesons prefer the interaction with air nuclei prior the decay. The very short lived $D$'s and $\Lambda_c$'s promptly decay into leptons, which carry a large fraction of the original energy of the nucleus. The model in qgsjet01c predicts that the purely prompt flux can be observed above 10 PeV for muons or neutrinos irrespective the direction of incidence.

\subsection{Differential lepton fluxes at the surface}
In contrast to the previously presented ratios, the total muon neutrino intensity can directly be observed by recent neutrino detectors, such as the IceCube Observatory and ANTARES. However, the uncertainty of the measurements is to some extent dependent on the model of hadronic interactions, the atmosphere and the primary flux assumed during the data analysis. Beside the cHGp primary flux parametrization usage throughout the calculations of the next subsections, we employ the GH spectrum when it is possible to compare the result of this study with measurements. In Sec.\ \ref{ssec:primaryvar} the influence of the primary model on the fluxes is studied explicitly. The neutrino fluxes are averaged over the azimuth and zenith angles, while the muon fluxes are presented for the vertical direction only. The prompt component is turned on in all simulations with qgsjet01c.

\begin{figure}[htb!]
\centering
\includegraphics[width=0.494\textwidth]{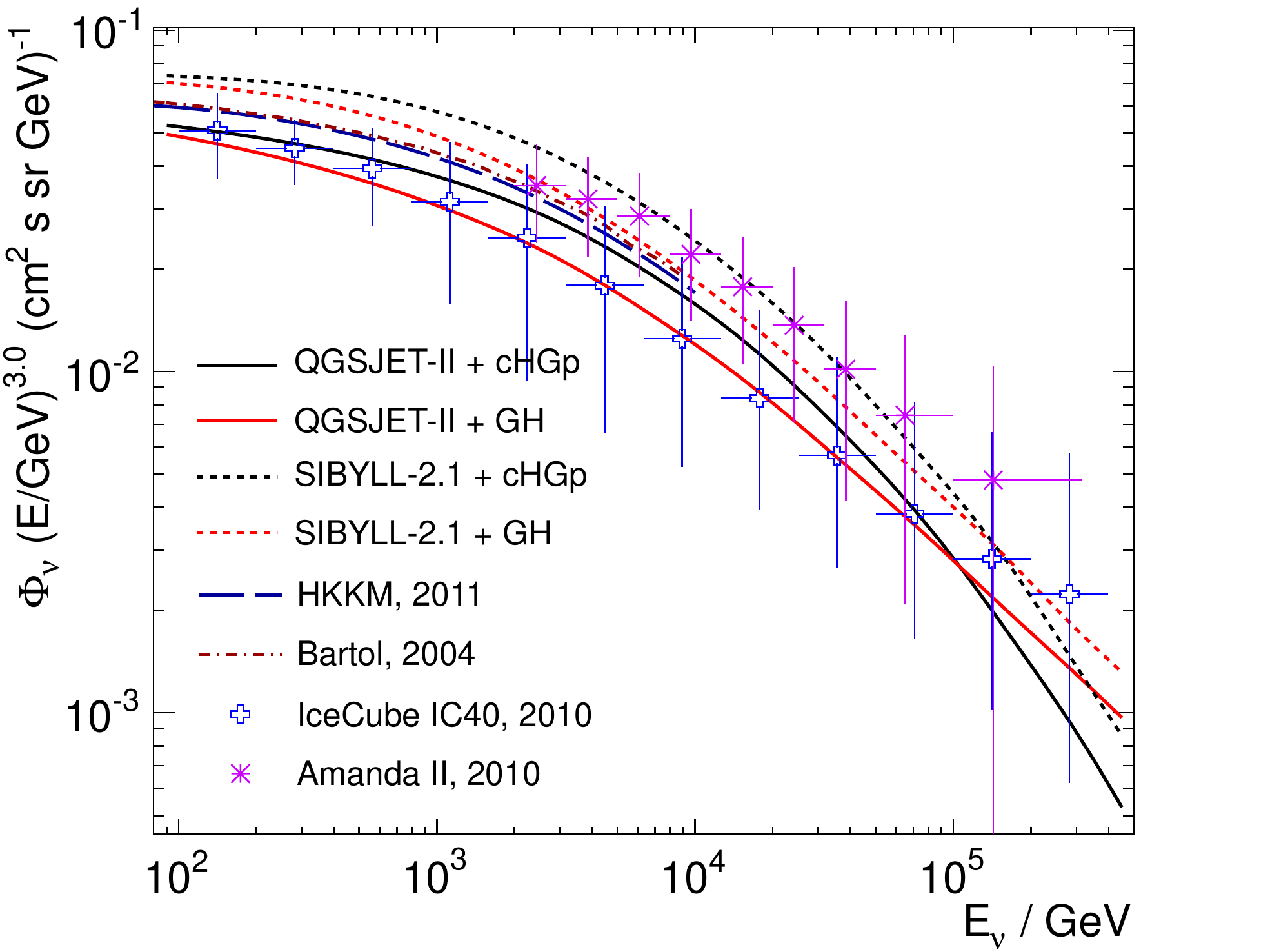}
\includegraphics[width=0.494\textwidth]{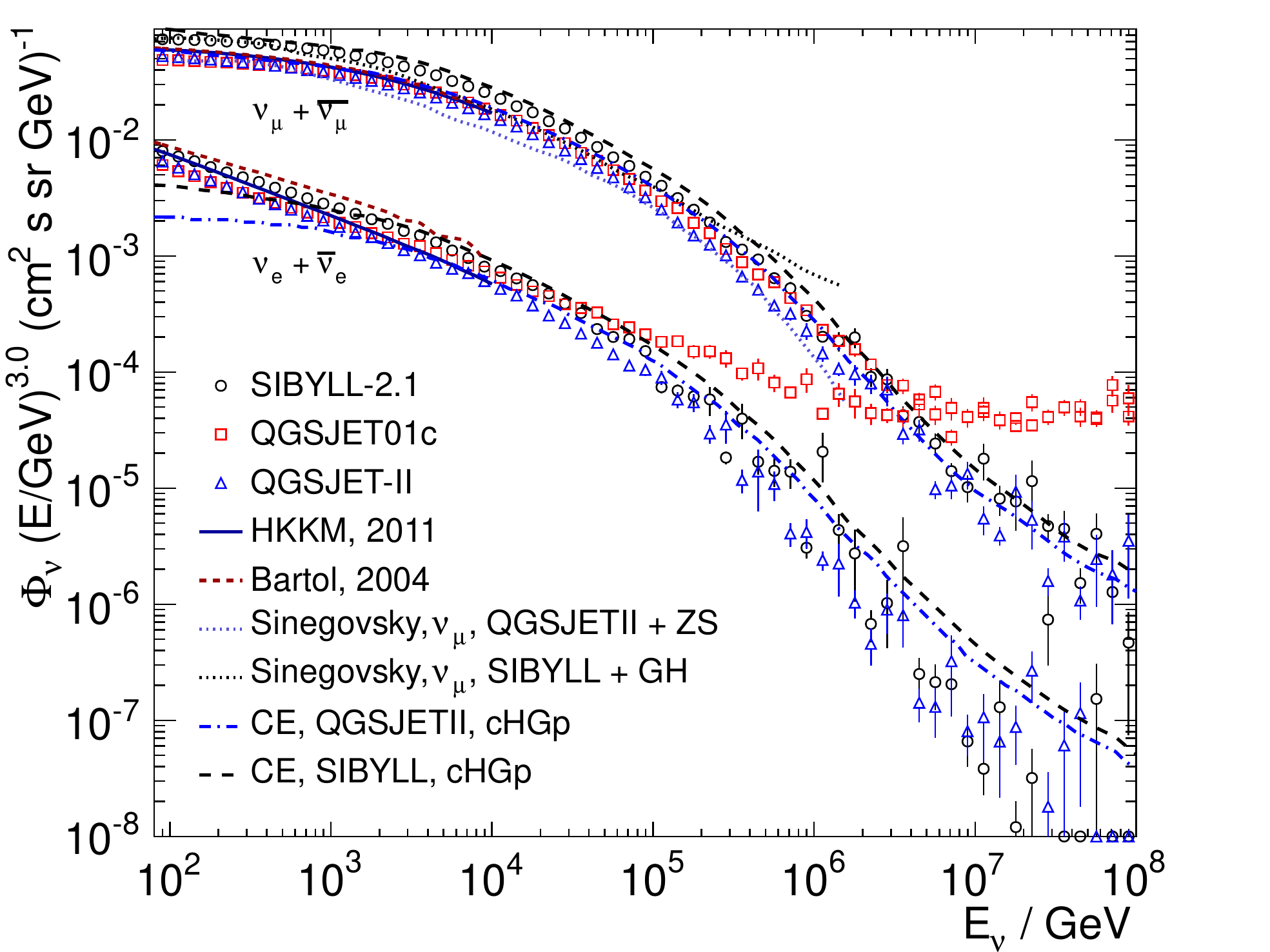} 
\caption{Flux of conventional $\nu_{\mu} + \bar{\nu}_{\mu}$ averaged over zenith and azimuth angles, compared to experimental data from Icecube in 40-string instrumentation \citep{icecube_ic40_2010} and Amanda II \citep{amanda_2010}. The bottom pane shows the full energy range for muon and electron neutrinos, calculated using the cHGp model and compared to calculations by \citep{sinegovsky_icrc_2011}, \citep{HKKM_2011}, \citep{bartol_2004} and our cascade equation approximation (CE).}
\label{fig:numuflux}
\end{figure}

The differential flux of conventional $\nu_\mu + \overline{\nu}_\mu$ is presented in Fig.\ \ref{fig:numuflux}, compared to experimental data from AMANDA II and IceCube with 40-string configuration. 

The flux obtained with the two qgsjet models is is comparable with Amanda and IceCube data in the experimentally observed range. As discussed in the previous section, {\sc sibyll} has a higher fraction of kaons leading to a higher muon neutrino flux. In connection with the harder primary spectrum compared to the previous calculations, it seems to overestimate the flux with respect to the IceCube data. In the region of hundreds of GeV, where the GH primary model is valid, the Monte-Carlo calculation agrees well with the semi-analytic calculations from \citet{sinegovsky_icrc_2011} and it is close to our semi-analytical approximation. 

The calculation of the flux of electron neutrinos is depicted in the bottom pane of in Fig.\ \ref{fig:numuflux} for the full energy range of the simulation. Due to the lack of experimental observations the results can only be compared to other calculations. The flux calculated with {\sc sibyll} or either version of qgsjet agrees with other calculations. Our semi-analytical approximation does not include muon decay, thus the results do not agree at lower energies. At the high-energy tail the statistics of the simulation seems insuffcient to predict the electron neutrino flux, or there are other not yet understood effects. The significant contribution of prompt neutrinos practically eliminates the effect of the knee on the spectral shape with qgsjet01c. Therefore, every single electron neutrino detected with $E_\nu > 1$ PeV has a very high probability to have a prompt or astrophysical origin. 

\begin{figure}[htb!]
\centering
\includegraphics[angle=90,width=0.494\textwidth]{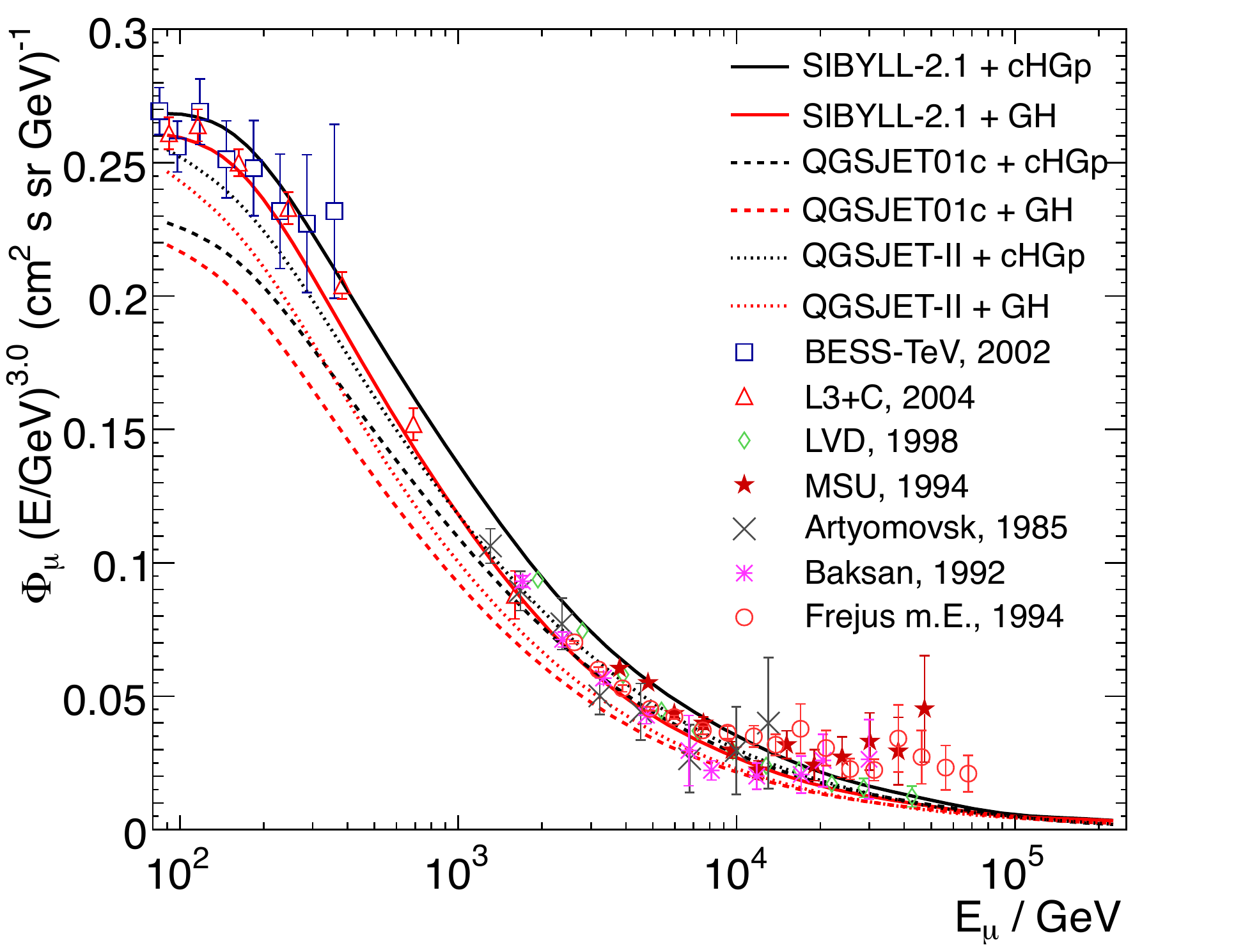}
\includegraphics[width=0.494\textwidth]{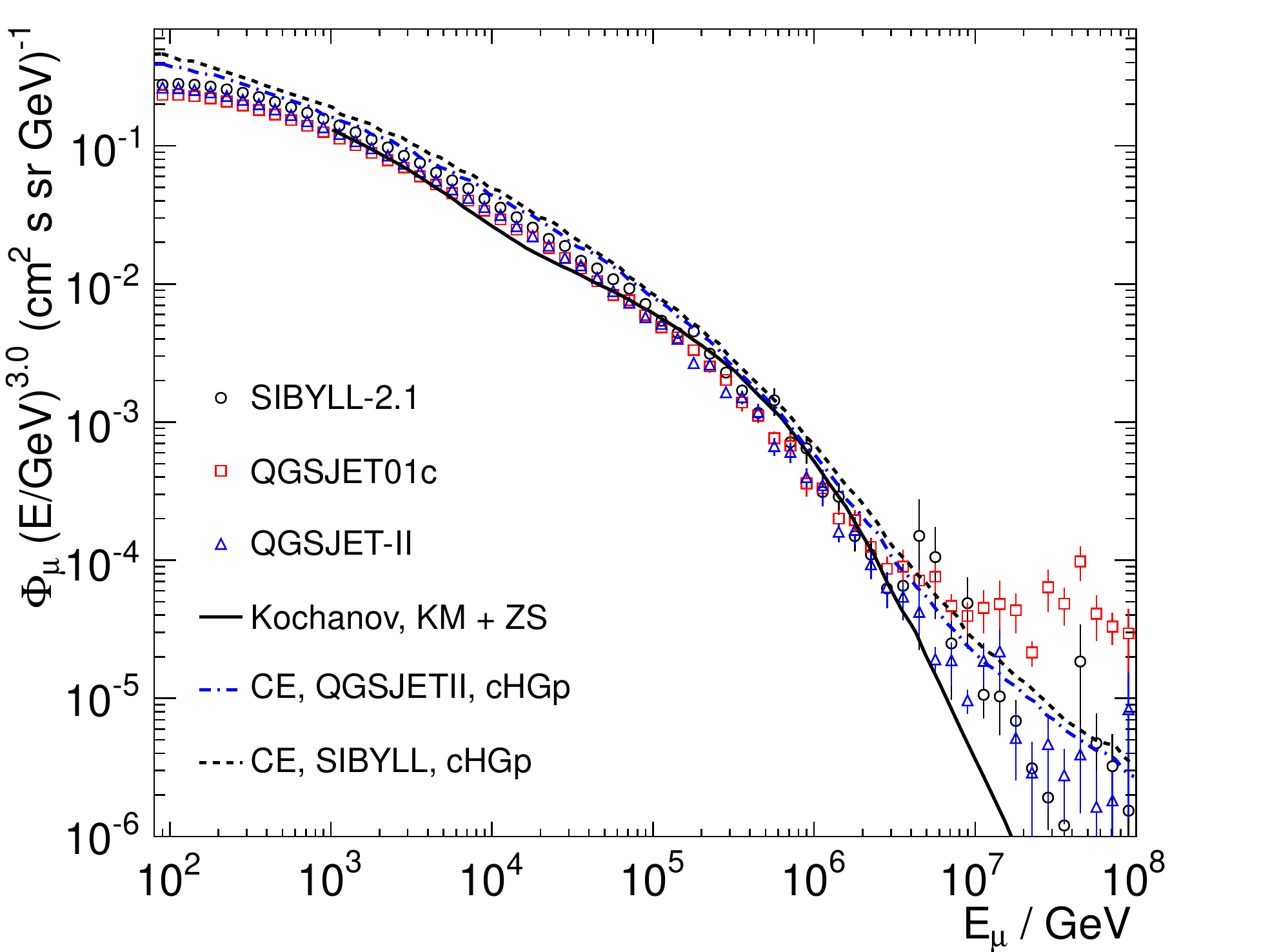}
\caption{Surface flux of vertical (on the left pane for $\cos(\theta) = 1.0$) $\mu^+ + \mu^-$ compared to measurements and calculations. The cascade equation calculations are from \citep{kochanov_2008}.  The experimental data are from \citep{artyomovsk_1985,baksan_1992,Frejus_1994,msu_1994,LVD_muflux,BESS_TeV,L3_C_muons} and datapoints for \citep{artyomovsk_1985,baksan_1992,Frejus_1994,msu_1994} have been taken from \citep{kochanov_2008}.}
\label{fig:muflux}
\end{figure}

In Fig.\ \ref{fig:muflux} the muon flux is compared to data and calculations.
The spread between the different interaction models decreases due to the minor contribution of kaons. In the range of the pion's critical energy ($\epsilon_\pi \approx 115$ GeV), the BESS and L3+C data constraints the validity of the interaction models. The calculation using {\sc sibyll} and either of the primary fluxes lies within experimental uncertainties. The case of {\sc {\sc qgsjet-ii}} + cHGp is at the lower error boundary of BESS-TeV data, while {\sc QGSJET-01} generally underestimates the differential muon spectrum. Because at higher energies the muons' energy at the surface has to be estimated from underground measurements, a direct comparison is not trivial.

\subsection{Variation of the primary cosmic-ray flux}
\label{ssec:primaryvar}
\begin{figure}[htbp]
	\centering
	\includegraphics[width=0.494\textwidth]{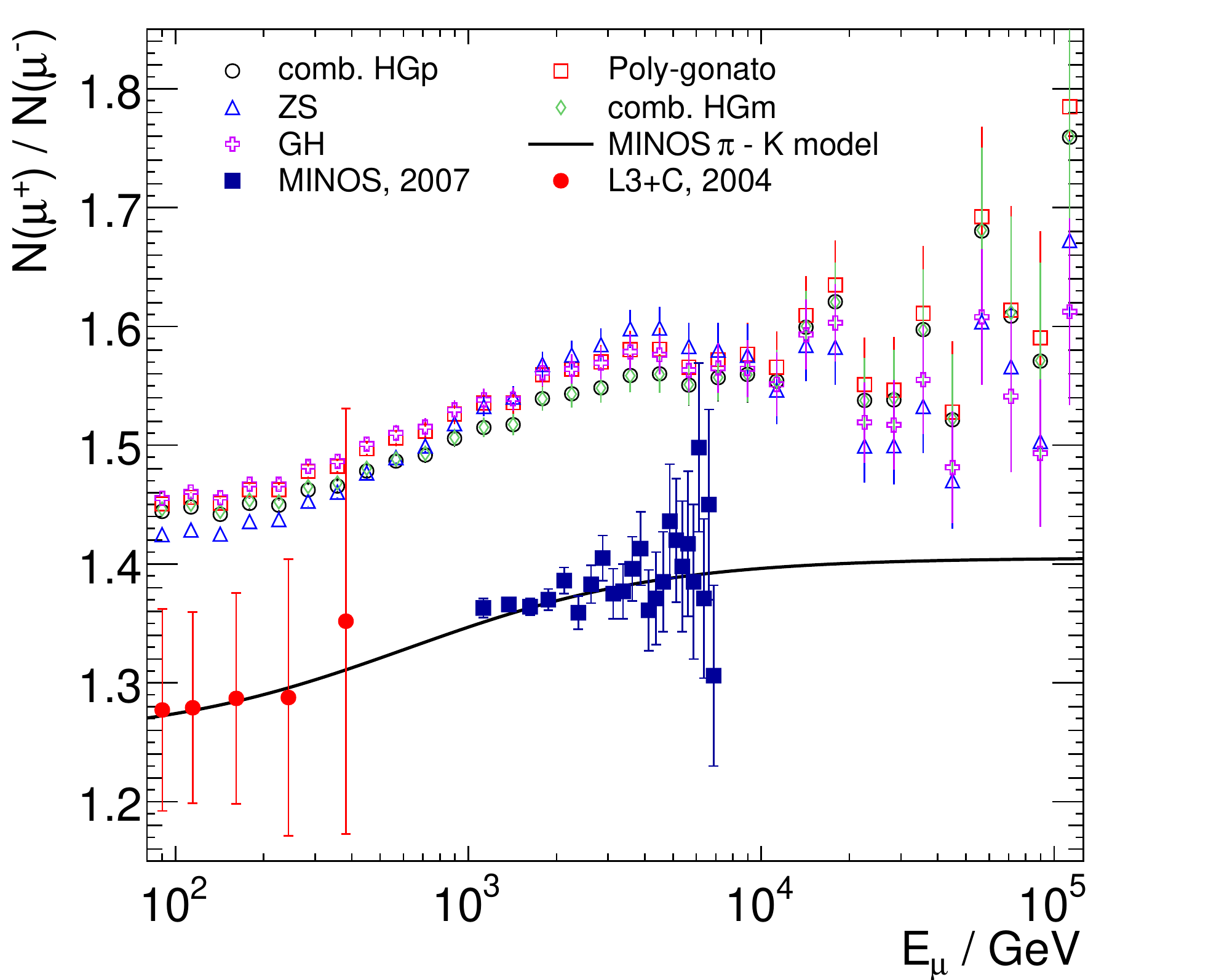} 
	\caption{The charge ratio, deduced from the vertical muon flux, using {\sc sibyll}-2.1 and different primary models. The MINOS $\pi$-K model and the data are from \cite{minos_mucharge} and for L3+C \cite{L3_C_muons} respectively.}
	\label{fig:mucharge_vs_primary}
\end{figure}

Figure \ref{fig:mucharge_vs_primary} shows the muon charge ratio, 
calculated with {\sc sibyll}-2.1 using different primary flux models. Although {\sc sibyll}'s prediction is too high, it does well reproduce the transition between the pion charge dominated (E $< \epsilon_\pi$) and the kaon charge dominated (E $>\epsilon_K$) regions of the charge ratio. As pointed out in \cite{tom_primary_muoncharge}, the variation of the primary flux model leads to the variation of the slope of this transition, which is steepest for the ZS spectrum and flattest for the two HG models. 

\begin{figure*}[htbp]
	\centering
	\includegraphics[width=0.495\textwidth]{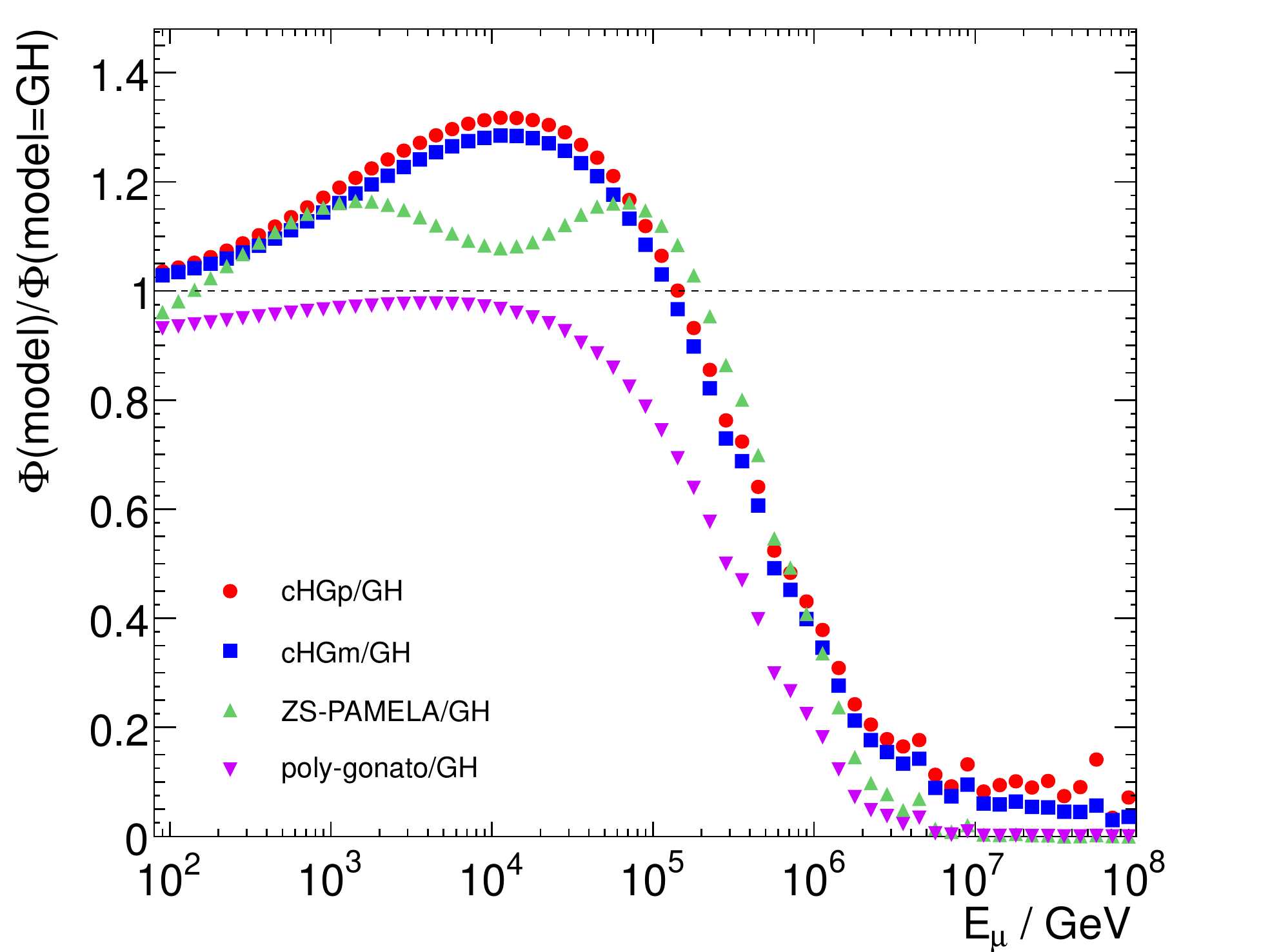}
	\includegraphics[width=0.495\textwidth]{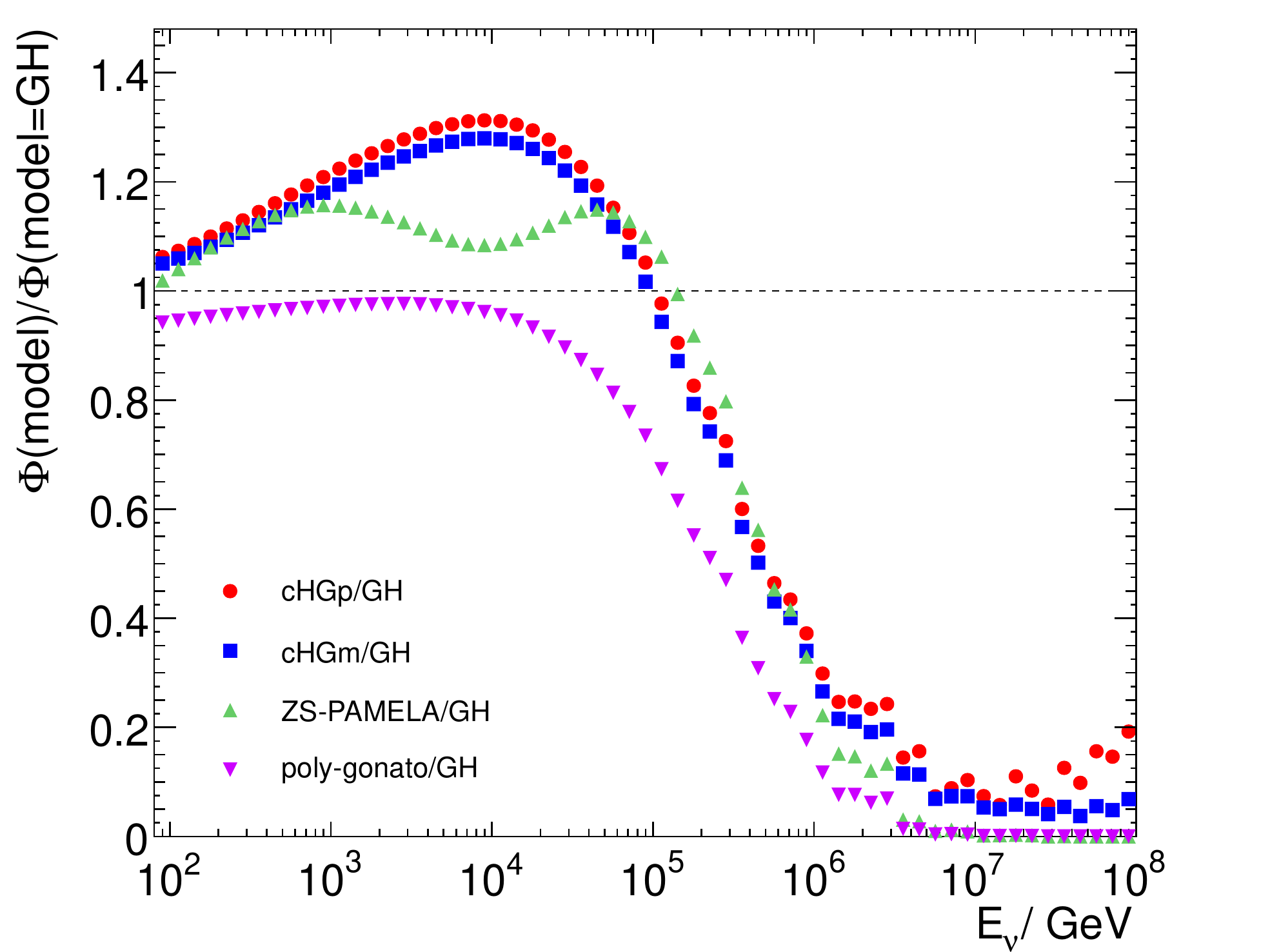} 
	\caption{Illustrated are the deviations of the muon (\textit{left}) and muon neutrino (\textit{right}) flux when calculated with different primary models, divided by the flux generated using the GH spectrum. The GH model has been extrapolated up to the highest energies, symbolizing a knee-less assumption. The selected interaction model is {\sc qgsjet-ii}.}
	\label{fig:primary_variation}
\end{figure*}

In Fig.\ \ref{fig:primary_variation}, 
we have calculated the surface fluxes assuming different primary spectra and compositions for the primary cosmic-ray flux with respect to a baseline spectrum. To emphasize the differences of this calculation in connection with previously published primary cosmic-ray flux models, we have selected GH (2002) as the baseline. The results are similar for {\sc qgsjet-01c}. The shape of these curves does not change when using {\sc sibyll}-2.1, but the features are shifted roughly a factor $\sim 2$ towards higher energies in the case of muon neutrinos and a factor of $\sim 4$  in case of muons, i.e.\ the ratio $\Phi_\mu(\mathcal{M} = \rm{cHGp})/\Phi_\mu(\mathcal{M} = \rm{GH})$ crosses unity at 800 TeV instead of 200 TeV. 

The poly-gonato model yields the lowest flux, falling below all other models above 500 GeV. This model is designed with the goal to describe the cosmic-ray flux below the knee and at the knee. Above, the spectrum is to steep and does not agree with data (see Fig.\ \ref{fig:primary_models}). It is therefore not suited to accurately describe effect of the knee on atmospheric leptons.

The Zatsepin-Sokolskaya (PAMELA parameters) model agrees with several indirect measurements at energies close to the knee and with direct PAMELA measurements in the proton and helium component. Using this model the lepton fluxes show a significant kink at tens of TeV, originating from the transition of the first (SN) to the second (SN into super-bubble) source class. This transition leads to a variation of the lepton fluxes in the order of 20 - 30\%. 

The two Hillas-Gaisser models (cHGp and cHGm) incorporate the hardest spectrum, and thus lead to the highest fluxes at lepton energies above several TeV. The hypothetic second Galactic component plays an important role at the knee, being the source of atmospheric leptons at knee energies. Due to the overall good agreement of these models with the all-particle primary spectrum we favor the version with protons in the extragalactic component for all our neutrino calculations.

The effect of the knee of cosmic-rays in the primary spectrum is reflected in a similar shape for muons and muon neutrinos. However, the logarithmic abscissa does not represent well the differences in energy. The spectral index begins to change at several TeV and falls below the knee-less hypothesis (GH spectrum) in the range $100$ - $200$ TeV in the case of the ZS and the cHGp/m models. 

\subsection{Charm in {\sc qgsjet-01c}}
\begin{figure}[htbp]
	\centering
	\includegraphics[width=0.495\textwidth]{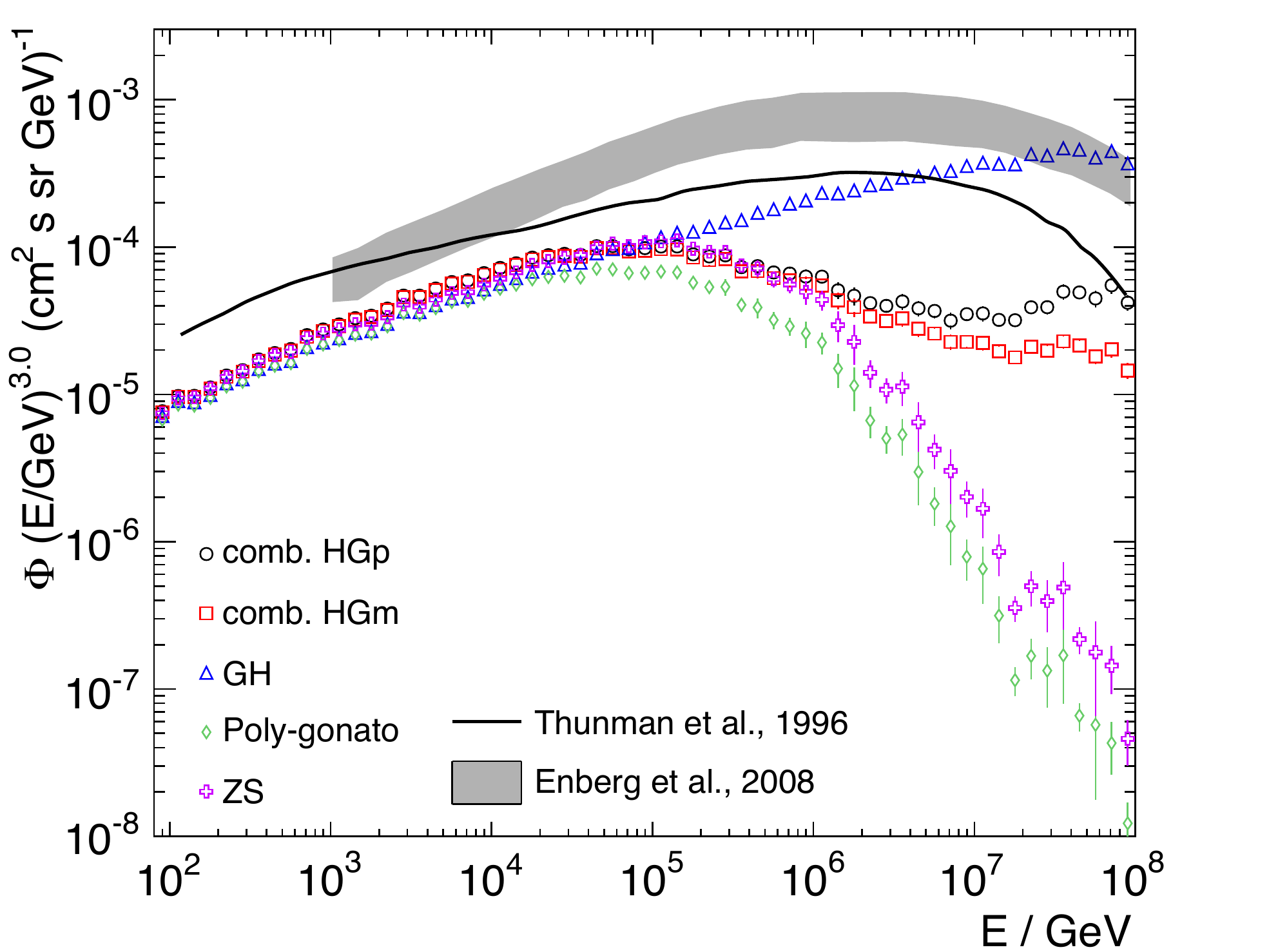}
	\caption{(Prompt) Flux of leptons originating from charm particles, calculated with {\sc qgsjet-01c}.}
	\label{fig:charm}
\end{figure}

We have used the implementation of charm hadron production in {\sc qgsjet-01c} to assess the influence of the primary spectrum and composition on the prompt flux. To minimize the statistical uncertainty the prompt flux has been averaged over the flavor ($\nu_\mu$, $\nu_e$ and $\mu$), since the cross-sections and branching ratios in the considered channels are nearly equal \cite{enberg_2008}. The results are shown in Fig.\ \ref{fig:charm}.
In general, the absolute flux generated with qgsjet is at least factor of $\sim 2$ lower in the range compared to the two reference calculations, the spectral index shows compatible characteristics up to energies of hundreds of TeV. Since the crossover between the conventional and the prompt flux occurs in a region above the knee, where the spectral index of the conventional flux has its maximum, the uncertainty is too large for a detailed discussion. Therefore, we restrict ourselves to the discussion of the influence of the primary model.

In the region below 100 TeV where the conventional fluxes dominate, the variation of the prompt component due to the primary model is smaller than the differences between the theoretical predictions in Fig.\ \ref{fig:charm}. Above the knee the influence of the cosmic-ray intensity and composition is evident. Primary models, which do not include a second Galactic or a third extragalactic component, suffer from a steep cutoff (poly-gonato and ZS). The differences between the cHGp and cHGm models show, that for the highest energies the modeling of an extragalactic component is crucial and that different composition scenarios of the primary flux influence the total rate of prompt particles.

\subsection{Estimation of the theoretical uncertainty}
\begin{figure}[htbp]
	\centering
	\includegraphics[width=0.495\textwidth]{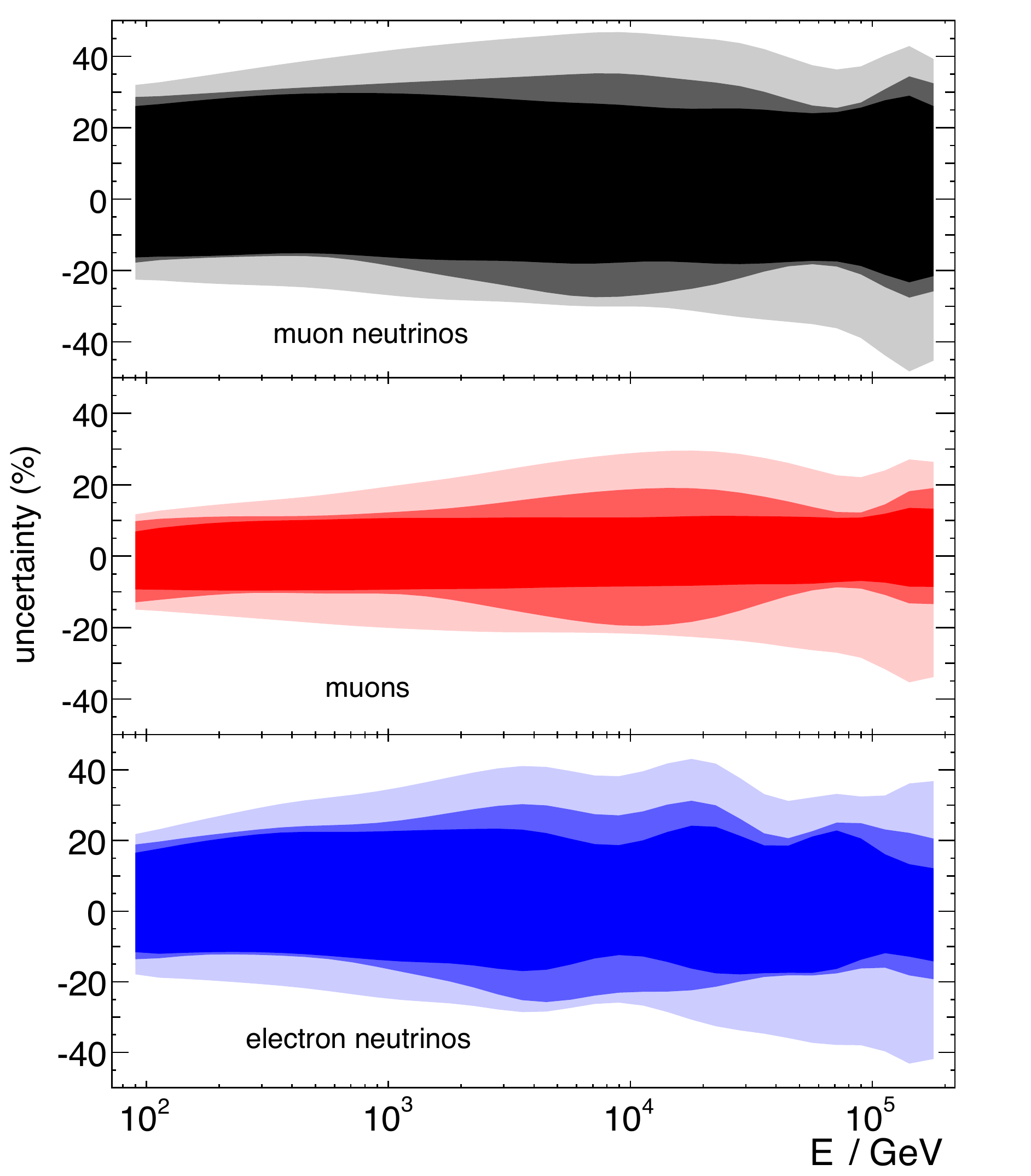}
	\caption{Total theoretical uncertainty derived from the variation of the interaction model is represented by the \textit{inner solid bands}. The shaded bands are derived from the variation of the interaction model using all available primary models (cHGp, cHGm, ZS/PAMELA for the \textit{medium shaded bands} and cHGp, cHGm, ZS/PAMELA, GH and poly-gonato for the \textit{outer shaded bands}).}
	\label{fig:ia_mod_unc}
\end{figure}

The total theoretical uncertainty of the conventional flux of atmospheric muons and neutrinos, derived from the results of this calculation, is shown in Fig.\ \ref{fig:ia_mod_unc} and Table ~\ref{table:conv_uncertainties}.
\begin{table*}
\caption{Theoretical uncertainties of the conventional atmospheric lepton fluxes, given in \%. {\it ALL} - variation of the combinations between the all primary models (cHGp, cHGm, ZS-PAMELA, poly-gonato and GH) and all hadronic interaction models {\it HIM} ({\sc sibyll}-2.1, qgsjet01c, {\sc qgsjet-ii}). {\it  RECENT} contains the same set as {\it ALL}, excluding the GH and the poly-gonato models. {\it HIM only} - primary model is fixed to cHGp and all interaction models are varied. The average values (av) are calculated using all data points from the bands.}
\vspace{0.5cm}
\begin{tabular}{| c | c | c | c | c | c | c | c | c | c | c | c | c | c | c | c |}
\hline \hline
 \multicolumn{1}{|c} {Type} & \multicolumn{5}{|c} {ALL + HIM} & \multicolumn{5}{|c} {RECENT + HIM} & \multicolumn{5}{|c|} {HIM only}\\
\hline
Energy (TeV) & $0.1$ & $1$ & $10$ & $100$ & av & $0.1$ & $1$ & $10$ & $100$ & av & $0.1$ & $1$ & $10$ & $100$ & av\\
\hline
\multirow{2}{*} {$\nu_\mu$} & $+32$ & $+42$ & $+47$ & $+39$ & $+42$ & $+29$ & $+32$ & $+35$ & $+29$ & $+32$ & $+26$ & $+30$ & $+26$ & $+27$ & $+27$\\
& $-23$ & $-27$ & $-30$ & $-41$ & $-32$ & $-17$ & $-19$ & $-27$ & $-23$ & $-22$ & $-16$ & $-16$ & $-18$ & $-20$ & $-20$ \\
\hline
\multirow{2}{*} {$\nu_e$} & $+23$ & $+35$ & $+39$ & $+33$ & $+35$ & $+19$ & $+25$ & $+28$ & $+24$ & $+25$ & $+17$ & $+23$ & $+19$ & $+18$ & $+20$\\
& $-18$ & $-25$ & $-26$ & $-39$ & $-30$ & $-13$ & $-16$ & $-23$ & $-16$ & $-19$ & $-12$ & $-14$ & $-13$ & $-13$ & $-13$ \\
\hline
\multirow{2}{*} {$\mu$} & $+12$ & $+20$ & $+29$ & $+23$ & $+24$ & $+10$ & $+12$ & $+19$ & $+13$ & $+15$ & $+7$ & $+11$ & $+11$ & $+11$ & $+11$\\
& $-15$ & $-20$ & $-22$ & $-30$ & $-24$ & $-13$ & $-11$ & $-19$ & $-10$ & $-13$ & $-9$ & $-9$ & $-8$ & $-7$ & $-7$ \\
\hline
\end{tabular}
\label{table:conv_uncertainties}
\end{table*}
The shaded bands were calculated using the average particle spectrum:
\begin{equation}
\langle \Phi_p(E) \rangle = \frac{1}{N_{\mathcal{M}} N_{\Phi_\mathcal{C}}}\sum_\mathcal{M}\sum_{\Phi_{\mathcal{C}}}\Phi_p(E,\mathcal{M},\Phi_{\mathcal{C}})
\end{equation}
The upper and lower boundaries of the uncertainty bands are then given by
\begin{equation}
\delta^+(E) = \max\frac{\Phi_p(E,\mathcal{M},\Phi_{\mathcal{C}})}{\langle \Phi_p(E) \rangle} \qquad
\delta^-(E) = \min\frac{\Phi_p(E,\mathcal{M},\Phi_{\mathcal{C}})}{\langle \Phi_p(E) \rangle}\,.
\end{equation}
This is for the combined uncertainty due to the interaction model and the primary flux model. Since the GH and the poly-gonato model do not consider a steepening and the disagreement of single power law fits to observations of direct measurements, we provide in the medium shaded band the uncertainty when using cHGp, cHGm and ZS as candidate spectra only. This result suggests that up to 1 TeV neutrino or muon energy, the uncertainty of the flux is dominated by the hadronic interaction model, while the flux of primary cosmic-rays and the composition ($< 100$ TeV/nucleus) are relatively well known. Above this energy the different assumptions about the origin and shape of the knee become dominant and result in additional 15\% to the uncertainties of the hadronic interaction models. At lepton energies approaching 100 TeV the uncertainties due to the primary flux are decreasing, since the primary nuclei responsible for these particles are from an energy range at the knee, where the primary models are compatible with each other.

The pure interaction model uncertainty was calculated by fixing the primary model $\Phi_{\mathcal{C}}$ to cHGp. The result shows the features discussed in Sec.\ \ref{ssec:iam_performance}. The weaker dependence on the representation of kaons in the air showers and a better overall agreement of the pion performance between the interaction models leads to the determination of the muon flux with a high precision ($< 11$ \%) up to hundreds of TeV. The important role of kaons for atmospheric neutrino production results in high uncertainties for both neutrino flavors. For the production of muon neutrinos, pions and the $K/\pi$-ratio play a bigger role, thus resulting in higher uncertainties for muon- than for electron neutrinos.
Our results for the interaction model uncertainty are somewhat higher when compared to the detailed study of uncertainties in atmospheric neutrino fluxes by \citet{barr}, which predicts for muon neutrinos at 1 TeV a total (gaussian) uncertainty of 30\% due to hadronic interactions and of 40\% if the primary flux is taken into account.

\subsubsection{Uncertainties due to the composition in the UHE regime}
\begin{figure}[htbp]
	\centering
	\includegraphics[width=0.495\textwidth]{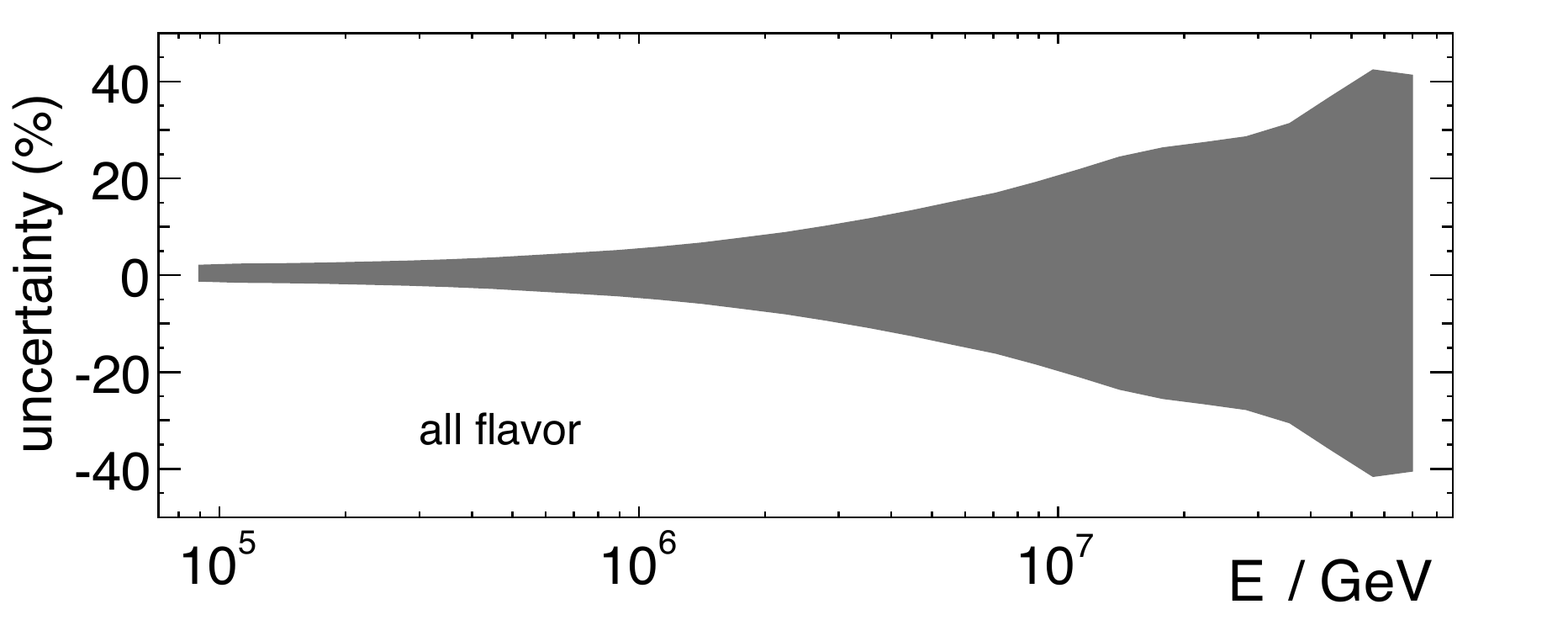}
	\caption{Theoretical uncertainty of the prompt atmospheric leptons due to the composition of the extragalactic component component.}
	\label{fig:iuhe_unc}
\end{figure}

Since the flux of muons and neutrinos at energies above hundreds of TeV is dominated by prompt muons and neutrinos, only {\sc qgsjet-01c} with an enabled charm component is suitable for studies. In Fig.\ \ref{fig:iuhe_unc}, 
the same approach as in the previous section has been chosen to identify the uncertainty of the UHE component due to the uncertain composition of extra-galactic cosmic-rays, resulting in less than $\pm 2$\% at 100 TeV and $\pm 40$ \% at hundreds of PeV.

\section{Summary and Discussion}
\label{sec:discussion}
We have developed a full Monte-Carlo calculation scheme which is capable to calculate muon neutrino, electron neutrino and muon fluxes up to 100 TeV, with a statical accuracy around a few percent. For surface energies up to 100 PeV, it was possible to assess the influence of the spectrum and composition of primary nuclei on the calculation. With respect to the increasing sensitivity of modern neutrino telescopes up to these energies, the influence of the knee of cosmic-rays has been studied from the perspective of lepton spectra at the surface.

The asymmetric uncertainties in the calculation of the fluxes have been assessed by carrying out the calculation using several models of the primary cosmic-ray flux and three interaction models, which individually represent different assumptions about hadronic interactions. It has been found that the uncertainties in the calculation of the atmospheric muon flux are significantly smaller, compared to the uncertainties of the atmospheric neutrino flux. As it has been shown in the study of flavor and charge ratios, this behavior can be explained by the insufficiently known contribution of kaons in the atmospheric cascade and the higher importance of kaons for the neutrino flux.

In particular, uncertainties of the conventional atmospheric neutrino flux are important as a dominant source of systematic uncertainties in the search for astrophysical high-energy neutrinos. Recent searches for different astrophysical neutrino signals with IceCube are based on the Gaisser-Honda parametrization of the spectrum with an assumed systematic uncertainty of 25\% \cite{icecube40_diffuse,icecube40_ps}. Here, it could be shown that depending on the energy, the uncertainty of the atmospheric neutrino flux is $^{+32}_{-22}$\% on average when using a realistic cosmic-ray spectrum. While this number is dominated by the interaction model, there is still some significant contribution from the primary flux models. If a better understanding of the composition and spectral behavior up to the knee and above can be achieved, the total uncertainty can be reduced further. In addition, it is expected that the inclusion of new LHC data can reduce the systematic uncertainties coming from the interaction models \cite{lhc_constraints} at the energy of the knee.

Using the charm option provided with {\sc qgsjet-01c}, it was possible to calculate the prompt component of atmospheric leptons. Although, the absolute value of the prompt flux is significantly lower than expected from other calculations, we were able to show the role of the primary flux model in this type of calculations. The prompt flux has been identified to be sensitive to the composition of the extragalactic cosmic-rays.

\begin{acknowledgments}
We are grateful to T.K.\ Gaisser and R.\ Engel for their support and their interest in this work. We would like to thank C.\ Baus, P.\ Berghaus, K.\ Brodatzki, X.\ Gonzales, F.\ Halzen, M.\ Olivo, W.\ Rhode, M.\ Unger and the entire IceCube collaboration for many helpful discussions and comments. We would also like to thank Dieter Heck for sharing his {\sc corsika} experience with us. JKB and AF further acknowledge the support from {\it Mercator Stiftung}, with a contributing grant within the {\it MERCUR} project An-2011-0075. This work was also supported by the Research Department of Plasmas with Complex Interactions. For computer simulations, we gratefully acknowledge the usage of the Computer-Cluster LiDOng (TU Dortmund) within the Universit\"atsallianz Metropole Ruhr (UAMR). PD acknowledges the support from the U.S.\ National Science Foundation-Office of Polar Programs.
\end{acknowledgments}

\bibliography{publications_atmospaper}

\end{document}